\documentclass[structabstract]{aa}  
\usepackage{graphicx}
\usepackage{url}
\usepackage{txfonts}
 \usepackage{amssymb}
 \usepackage{natbib}
  \usepackage{lscape}
  \bibpunct{(}{)}{;}{a}{,}{,}
  \usepackage{times}
  \usepackage{float}
\usepackage[caption = false]{subfig}

  \def\deg{${}^\circ$}
  \def\min{${}^{\prime}$}
  
  \usepackage{longtable}
\usepackage{longtable,lscape}
  \usepackage{lscape}
\begin{document}
\def\hii{\ion{H}{ii} }
   \title{Low mass star formation and subclustering 
    in the \hii  regions RCW 32, 33 and 27
    of the  Vela Molecular Ridge}
   \subtitle{A photometric diagnostics to identify M-type stars}

   \author{L. Prisinzano\inst{1}
          \and
          F. Damiani\inst{1}
          \and
          M. G. Guarcello\inst{1}
           \and
          G. Micela\inst{1}
          \and
          S. Sciortino\inst{1}
          \and
          E. Tognelli\inst{2}
                    \and
          L. Venuti\inst{1}
}
   \institute{INAF - Osservatorio Astronomico di Palermo, Piazza del Parlamento 1, 90134, Palermo, Italy \\
              \email{loredana.prisinzano@inaf.it}
 	\and
 	Department of Physics `E. Fermi', University of Pisa, Largo Bruno Pontecorvo 3, I-56127 Pisa, Italy
         }
   \date{ }
 
  \abstract
   {Most stars born in clusters and recent results suggest that star formation (SF) 
   preferentially occurs in subclusters. Studying the morphology and SF history
   of young clusters is crucial to  understanding early cluster formation processes. 
 }   
   {We aim to identify the embedded population of young stellar objects (YSOs) down to the low mass stars in the M-type regime,
   in the three \hii regions RCW\,33, RCW\,32 and RCW\,27 located in 
    the northwestern region of the Vela Molecular Ridge. 
    Our aim is to characterise their
   properties, such as  morphology and extent of the clusters in the three \hii regions,
   derive stellar ages  and the connection of the SF
   history with the environment.
         }
  {Through  public photometric surveys such as Gaia, VPHAS, 2MASS and Spitzer/GLIMPSE, we identify YSOs with 
  classical techniques aimed at  detecting IR, H$\alpha$  and UV excesses, as signature of circumstellar disks
  and accretion. In addition, we implement a method to distinguish main sequence (MS) stars and giants in the M-type regime,
  by comparing the reddening derived in several optical/IR color-color diagrams, assuming suitable theoretical models.
  Since this diagnostic is sensitive to  stellar gravity, the procedure allows us to identify also
  pre-main sequence (PMS) stars.
  }
   {Using the classical membership criteria, we find a large population of YSOs
  showing 
   signatures of circumstellar disks with or without accretion. In addition, with the new technique of
    M-type  star
   selection, we find a rich population of  young M-type stars  with a spatial 
   distribution strongly correlated to the more massive population. We find evidence of three young clusters, 
   with different morphology, for which
   we estimate the individual distances by using TGAS Gaia
   data of the  brighter subsample. 
     In addition, we 
   identify field stars falling in the same region, by securely classifying them as giants and foreground MS stars.}
   {We identify the embedded population of YSOs, down to about 0.1\,M$_\odot$, 
   associated with the three \hii regions RCW\,33, RCW\,32 and RCW\,27 and the 
    three clusters Vela T2, Cr\,197 and Vela T1, respectively. All  the three clusters are located at a similar distance but
    show very different morphologies. Our results suggest a decreasing SF rate in Vela T2 
    and triggered SF in Cr\,197 and Vela T1.}

   \keywords{techniques: photometric -- stars: formation -- pre-main sequence -- stars: low mass, ISM: \hii  regions 
   Galaxy: open clusters and associations: individual: RCW\,33, RCW\,32, RCW\,27, Vela T1, Vela T2, Cr 291, stars: formation
               }
\authorrunning{L. Prisinzano et al.}
\titlerunning{Star formation in the Vela Molecular Ridge}

   \maketitle
%

\section{Introduction}
The Vela Molecular Ridge (VMR) is a large molecular cloud complex,
revealed by four strong emission peaks   observed in the 1$\to$0 transition
of $^{12}$CO \citep{murp91},
indicating that active SF is taking place \citep{pett08}.
It is located in the constellations of Puppis and Vela and it is part of a complex region 
including the very large \ion{H}{ii}  Gum Nebula, the Vela supernova remnant and also a 
remarkable annular system of cometary globules  \citep{pett08}.

\citet{murp91} identified four main clouds within the VMR, named A, B, C and D,
with the clouds A, C and D located at 0.7$\pm0.2$\,kpc and the B cloud located a 2\,kpc.
Several small \hii  regions have also been found in this region.
\citet{lise92} and \citet{lore93} have found 33 class\,I objects in the VMR
based on JHKL NIR photometry, of which 25 fall in clouds A, C and D  and 8 in the cloud B. 
Other surveys based on  NIR and CO observations have been conducted in the whole VMR region with the aim of finding signs 
of embedded YSOs. A detailed review is given in \citet{pett08}. 

The northwestern region  of the VMR, falling in  cloud D
 has been the subject of an H$\alpha$ survey
covering a 5\fdg5 $\times$ 5\fdg5 field, centered on 
RA (1950)=8$^{\rm h}$43\fm2, Dec  (1950)=-40\deg57\min by \citet{pett94}.
The presence of three \hii  regions, RCW 27, 32 and 33,
where signatures of recent low-mass SF
have been found by these authors,
is one of the peculiarities of this field. 
 In fact, they identified 278   H$\alpha$-emission objects (with a completeness limit V=19)
associated with dark clouds and the three \hii  regions. In particular, they found two major 
concentrations, called Vela T1 and  Vela T2 T Tauri associations, towards the \hii  regions RCW\,27  
and RCW\,33, respectively.
They took also slit spectra of 24 of the emission-line stars
    and found that the majority have a spectral appearance similar to T Tauri stars.
    
  \citet{yama99}   attempted to assign a possible age sequence,
    suggesting that  the sparse Vela T1 association, in RCW\,27, is  more evolved than the 
    tighter Vela T2 association in RCW\,33. They explain this sequential SF
    as an effect of the influence of the exciting  OB stars on the molecular clouds.

\citet{mass99,mass03} 
have carried out deep JHK imaging and 1.3 mm continuum photometry 
of 15 of the Class I sources found in the VMR D, while \citet{test01}  and \citet{mass06} have obtained deep JHKs images
 of six of the most luminous  IRAS sources  in the VMR D cloud.
They found evidences of embedded clusters younger than 1\,Myr for which they derived  
 Initial Mass Functions that are consistent with those derived for field stars and clusters.
  
 RCW\,32 is likely excited by  HD 74804, the brightest star of Collinder\,197 (Cr\,197).
 The H$\alpha$ emission stars in this region are very concentrated.  
 It  has been the target of a deeper objective prism image and of a ROSAT HRI observation,
 from which 70 and 30 YSOs, respectively, have been identified. Several embedded sources were found 
 in the region suggesting an age of 1\,Myr.
 The young galactic  cluster Cr\,197  has been the subject of a dynamical
 study by \citet{bona10}, who found it in a super-virial state, with evidence of deviation from dynamical equilibrium.
   
 Several studies, dedicated to individual sources lying in the northern region of the VMR,  provide
  significantly different distances in the range 0.70-1.15\,kpc \citep[see][for a review]{pett94}. 
 More recently, a distance of 0.85\,kpc has been attributed to the   cluster, named ASCC\,50,
 detected in  RCW\,33, by \citet{khar05} and confirmed by \citet{khar13}.
 
 In this work, we want to study the northern region of the VMR covered by \citet{pett94}
 to understand the connection between the star forming sites 
found in it,  and the SF history. In particular, we aim at characterizing and comparing the properties of
the three embedded young clusters Vela T2 (corresponding to ASCC\,50),
 Cr\,197 and Vela T1, associated with RCW\,33, RCW\,32 and RCW\,27, respectively.

To this aim, we exploit the potential of recently public  data as the deep optical-H$\alpha$ VPHAS survey
and the available NIR and X-ray data  complemented by the new Gaia astrometric and kinematic data.
This will allow us to perform a global analysis of the region, by assessing the  properties
of the three sub-clusters from which to derive hints on the SF process.

The paper is organized as follows. Sect. 2 describes the photometric catalog
 we assembled and used for this study. 
 In Sect. 3, we describe the selection of accreting YSOs while, in Sect. 4, we
 describe the selection of YSOs with a circumstellar disk. 
 In Sect. 5, we describe the kinematic analysis of the TGAS Gaia data, while in Sect. 6,
 we summarize the properties  of the members selected with classical methods.
 In Sect. 7 
 we illustrate the procedure adopted to identify M-type stars,
  derive their reddening, and  identify M-type giants, MS stars and cluster members.
   In Sect. 8, we discuss  our results on the spatial distribution and the SF history,
   while
    our conclusions are summarized in Sect. 9.



\section{Catalogue compilation}
\begin{figure*}
 \centering
    \includegraphics[width=15cm]{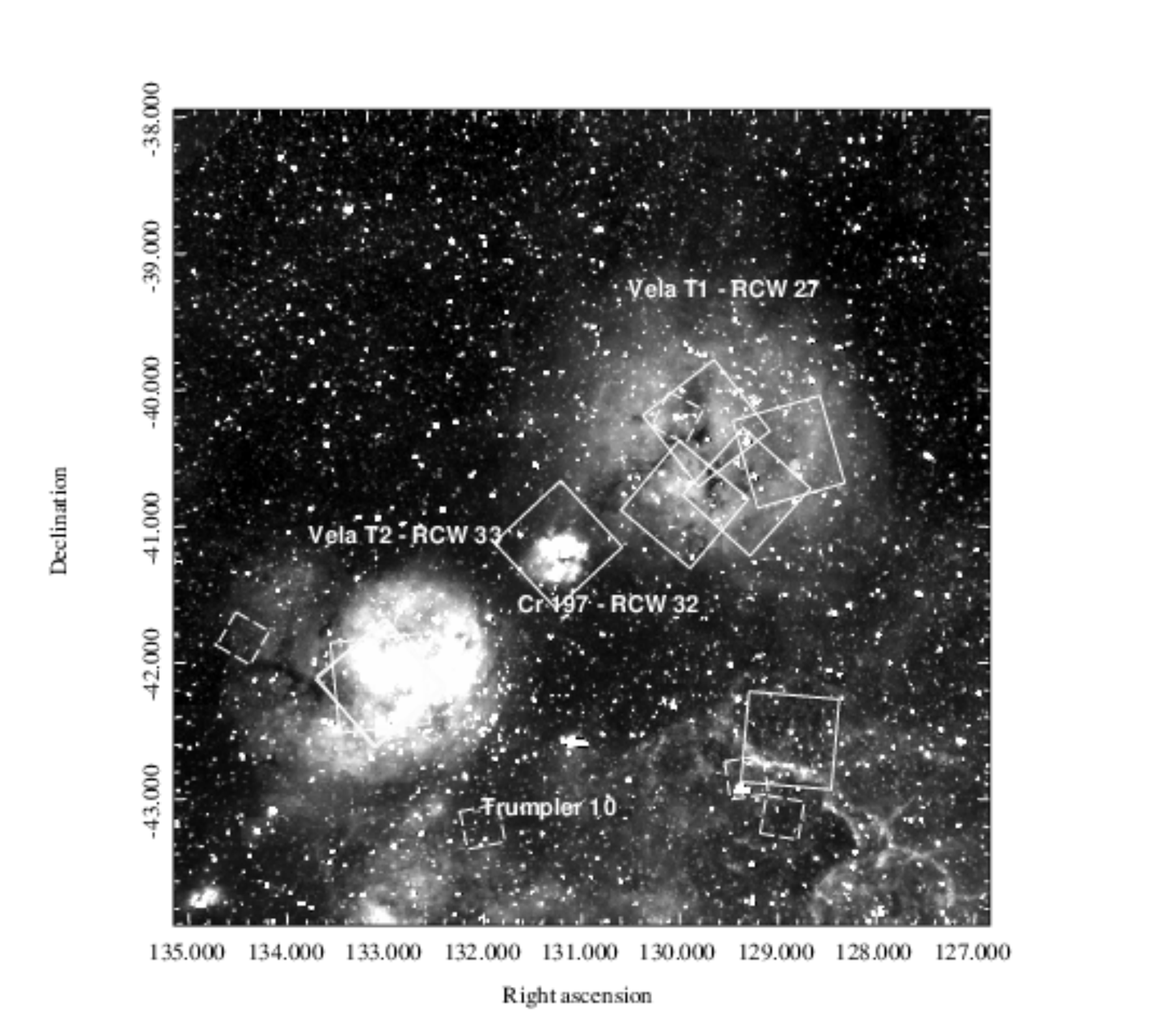}
\caption{A 6\deg $\times$ 6\deg\,\,H$\alpha$ image, centered on (RA, Dec)=(131, -41),
 taken from the Southern H-Alpha Sky Survey Atlas  \citep{gaus01}.
The positions of Trumpler\,10 and of the three \hii regions   are  indicated. The field of view  (FOV) of the 
ROSAT HRI and Chandra ACIS-I observations are indicated by solid and dashed boxes, respectively.}
\label{vrmhalpha}
 \end{figure*}

Figure\,\ref{vrmhalpha} shows an H$\alpha$ image of 6\deg$\times$6\deg  centered at 
(RA, Dec)=(131\deg, -41\deg) 
taken from the Southern H-Alpha Sky Survey Atlas \citep{gaus01}.
It includes  the northwestern region  of the 
VMR  we study here 
that we label  as  NW-VMR region. 
The three areas in the
three \hii  regions RCW\,33, RCW\,32 and RCW\,27, centered, respectively, on 
RA (J2000)=[132\fdg73,131\fdg2,129\fdg5] deg and Dec (J2000)=[-42\fdg13,-41\fdg3,-40\fdg6]  will be
 the subject of this work.

The NW-VMR is included in the VST Photometric H$\alpha$ Survey of the Southern Galactic
 Plane \citep[VPHAS+][]{drew14}.
We used photometry of Data Release 2 (Date: 2015-07-30)  from the ESO Catalogue Facility 
within a radius of 3\deg
around (RA, Dec)=(131\deg, -41\deg).
This multi-band source catalogue has been obtained from observations of 
the OmegaCAM CCD imager mounted on the 2.6 m VLT Survey Telescope (VST) on Cerro Paranal, Chile.
Imaging is performed with the IMAGE/OFFSET technique by using the bands	u\_SDSS, g\_SDSS, r\_SDSS, i\_SDSS, NB\_659,
corresponding to the five filters {\it u, g, r, i} and H$\alpha$, down to $r\sim21$.

 From this catalog, we selected objects tagged as
the  best available detections per unique object,
with  signal-to-noise ratio $>$ 10 and  a DAOPHOT \citep{stet87} point spread function fitting score of CHI $<$ 1.5 
in the r, i and H$\alpha$ bands.
 With this selection, 
the VPHAS+ catalog in NW-VMR includes 499\,664 objects with photometric errors
smaller than 0.1\,mag. 
Since the observations with the u and g filters cover only the northwestern  part
of  RCW\,27, we did not consider in this work the u and g magnitudes. 

VPHAS magnitudes are given both in the Vega and AB system   \citep{oke83}. 
A detailed description of the survey strategy,
  photometric offsets,  photometric quality,  exposure times and  pipeline used to 
 derive the magnitudes can be found in \citet{drew14}.
 
 The deep VPHAS+ photometric catalog ranges from  g=13 down to g$\sim$21.
Magnitudes g$<$13 are affected by saturation \citep{drew14}. 
For this reason, we compiled the photometric catalog
of the NW-VMR region by using  the  VPHAS+ catalog for r $\ge$13 and the 
AAVSO\footnote{American Association of Variable Star Observers.}
  Photometric All-Sky Survey (APASS) DR9 \citep{hend16}  catalog for r $<13$.
 To check the VPHAS+ calibration, we cross-matched the  APASS magnitudes, calibrated in the SDSS\footnote{Sloan Digital Sky Survey} 
   AB system,
   with the VPHAS+ catalog 
  by considering objects with $14<r<17$, both in the  VPHAS+ and APASS catalog. These limits have been chosen
  to avoid saturated objects in the VPHAS+ catalog and large errors in the APASS catalog whose limiting
  magnitude is about r=17.8 mag.
  By using a matching radius of 1\arcsec, we computed the systematic shift between the astrometric systems
   of the two catalogs corresponding to RA$_{VPHAS}$-RA$_{APASS}$=0\farcs07 and Dec$_{VPHAS}$-Dec$_{APASS}$=0\farcs05.
    We corrected the APASS coordinates for this shift
  and cross-correlated again the two catalogs by adopting a matching radius of 0\farcs5. 
 The 5\,982 
   objects common to VPHAS+ and APASS catalogs were used to check the VPHAS+ photometry.
  By considering the range 0.2$<r-i<1.5$, comprising the bulk of objects of our interest,
  we find that the r and i magnitude differences between  VPHAS+ and   APASS catalogs do not show any dependence
  on the relevant colors, while the median offsets  are
  0.009 and  -0.020 mag, respectively in the i and r bands. Since these  are 
  negligible with respect to the photometric errors,
   we do not apply any correction to the magnitudes.
   The final optical catalog includes 568\,811 entries in NW-VMR.

 The NIR counterparts of the objects included in our optical catalog 
were found in the 2MASS Point Sources  catalog   \citep{cutr03}  from which  we selected only objects 
with photometric quality flag 'A' in at least one of the three magnitudes JHK.
The cross-match between the two catalogs has been performed 
 by using 
the match service available at CDS, Strasbourg.
By adopting a matching radius of 1\arcsec, we found that 431\,860 of the  568\,811 entries in our
catalog have at least one counterpart in the 2MASS catalog. 

The field we are studying has been also included in the Vela-Carina Spitzer IRAC program,  an extension program of the
Galactic Legacy Infrared Midplane Survey Extraordinaire 
 \citep[GLIMPSE,][]{benj03,chur09}.
We used the  Vela-Carina Catalog \citep{zaso09},
that in NW-VMR includes 
170\,206 objects with IRAC/Spitzer  magnitudes.
We cross-matched the list of IRAC sources with  our  catalog by adopting a matching 
radius of 0\farcs5 and 
found that 
89\,263 IRAC sources have a counterpart in our optical catalog\footnote{the GLIMPSE point source accuracy is typically 0.3 arcsec}. 

In addition, we used the TGAS subset of the {\it Gaia} DR1 catalog \citep{gaia16,gaia16a},
limited to the objects included in the Hipparcos and Tycho-2 Catalogues,  for which we have
positions, proper motions  (PM) and parallaxes. 
The TGAS catalog was cross-matched\footnote{Using the match service provided by CDS, Strasbourg}
 with the APASS catalog 
in NW-VMR, finding 3\,065 matches within 1\arcsec.

In NW-VMR several X-ray sources were also found from the ROSAT High Resolution Imager  
(HRI, 1RXH, 3rd Release) and from the Chandra ACIS observations  taken from the Chandra Data Archive
 \citep{wang16}. The FOV of these observations are shown in Fig.\,\ref{vrmhalpha}.
We cross-matched the two X-ray catalogs,  including 25 Chandra  and 75 ROSAT sources (with 3 sources in common to  
the two catalogs),
 with our optical catalog by using the match service provided by CDS,
and adopting a radius of 5\arcsec. 
Within the NW-VMR area considered here,  we considered the   optical counterparts closest to the
97 X-ray   sources as YSOs.  

In particular, we have  8 and  34 X-ray sources from the ROSAT catalog
falling, respectively, in  RCW\,33 and RCW\,32, and  40 X-ray sources
in  RCW\,27, of which  36 from the ROSAT catalog and  7 from the Chandra ACIS catalog,
with 3 objects detected  with both Chandra and ROSAT.
The very small number of X-ray detections is due to the limited ROSAT sensitivity and to the small
exposure times (from 1 to 46\,ks) 
and coverage of the Chandra 
observations in this region.

\section{Candidate young stars with ongoing accretion from H$\alpha$ excesses\label{halphaexcesses}}
 CTTS stars are  YSOs with ages younger than few Myrs,  characterized by 
a circumstellar disk that transfers its matter onto the central star.
This process involves high velocities of the gas   impacting on the star
with a subsequent formation of a surface shock emitting predominantly in the UV, while 
H$\alpha$ emission is produced in the accretion funnel.
The first gives rise to the UV flux excess distinctive of CTTS, while the second
can be  photometrically detected  through the (r-H$\alpha$) color, that is a
measurement of the strength of the H$\alpha$ emission line relative to the photospheric continuum
in the r band. 

\begin{figure}
 \centering
 \includegraphics[width=9.5cm]{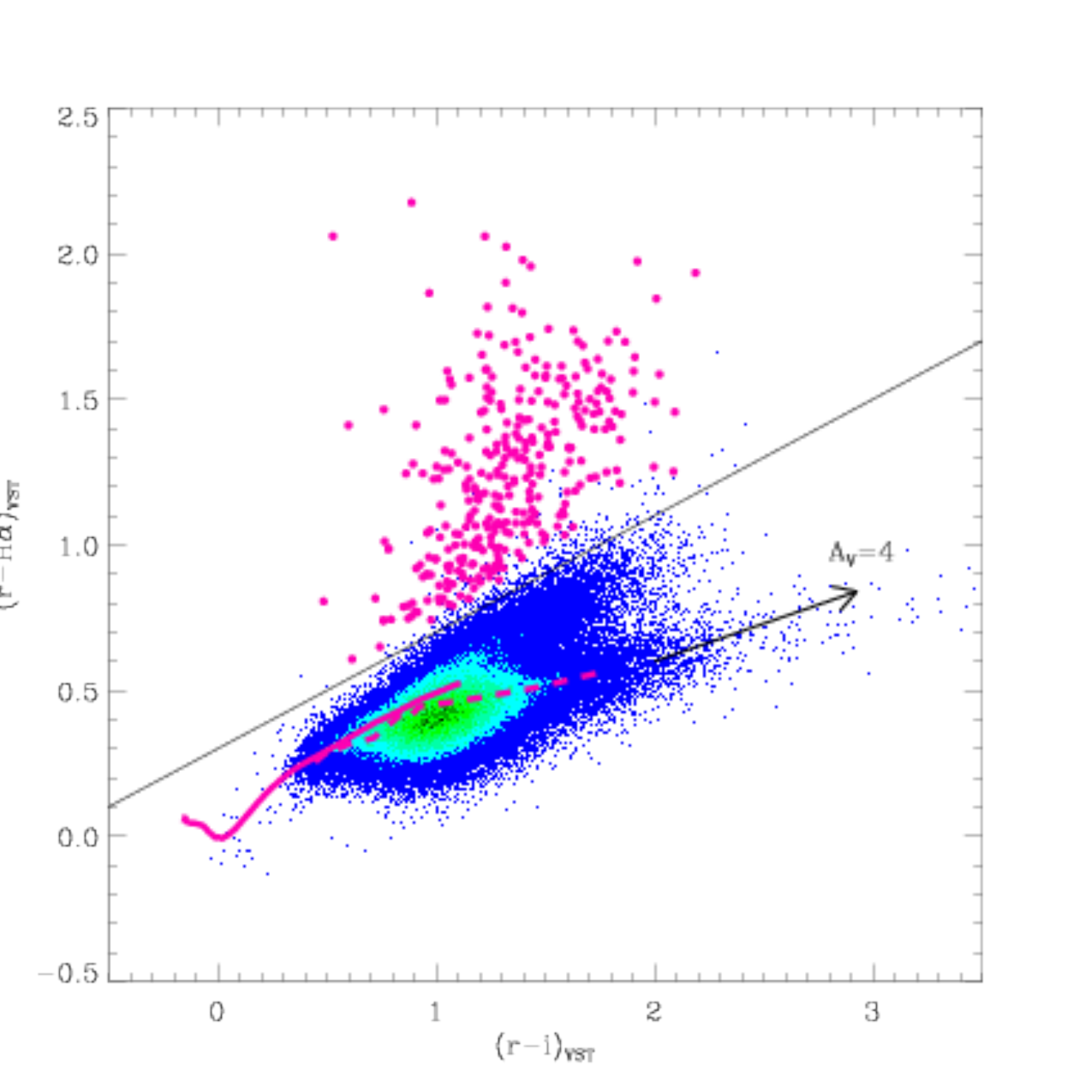}
\caption{VST r-H$\alpha$ vs. r-i diagram plotted as a two-dimensional stellar-density histogram 
using a binning of 0.005$\times$0.005 and a 8 level colour-coded scale such that high source densities 
(130 
  per bin) are dark-green and the lowest densities (one per bin) are violet.
  The magenta solid and dashed lines are, respectively, 
 the synthetic unreddened main sequence and giant loci given by \citet{drew14}, while the black solid line is 
 the limit adopted to select objects with H$\alpha$ excesses, drawn as magenta circles. }
\label{rirha}
 \end{figure}

Figure\,\ref{rirha} shows the  r-H$\alpha$ vs. r-i    color-color diagram (CCD) ,
 for all  473\,998 VPHAS+ objects  in  NW-VMR region 
 with color errors $<0.1$\,mag.
The synthetic unreddened main sequence for stars with spectral type from O6V to M2V  
and the giant locus for stars with spectral type from K0III to M5III,
computed by \citet{drew14}, 
are also shown. Since the NW-VMR region includes a very large area with patterns of dark obscuring matter,
 spatially variable  reddening is expected. This explains the large spread in  colors around the synthetic
loci. Neverthless, the bulk of main sequence and giant stars are enclosed within the blue region, while objects 
with larger r-H$\alpha$
 are stars with H$\alpha$ emission lines. We thus consider as stars with H$\alpha$ excesses those 
with  r-H$\alpha$ color larger (with at least 5$\sigma$ confidence) than the limit 
 r-H$\alpha$=0.3+0.4$\times$ (r-i).
In this way we select 329 CTTS candidates.

  \begin{figure}[!h]
 \centering
 \includegraphics[width=9.5cm]{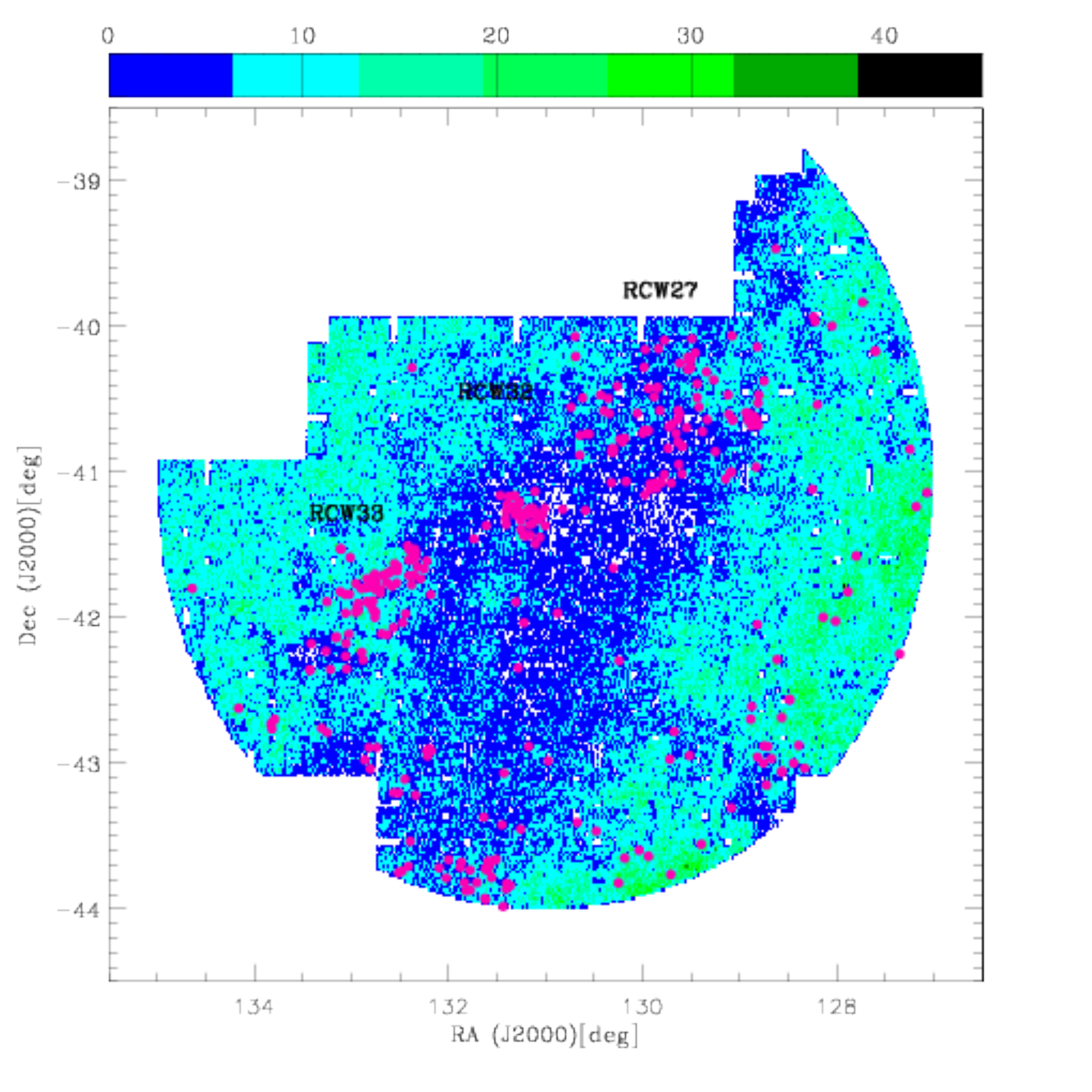}
\caption{Spatial distribution of the VPHAS+ stars in the NW-VMR region,  plotted as a two-dimensional stellar-density histogram 
using a binning of 0\fdg02$\times$0\fdg02 and a 7 level colour-coded scale such that high source densities 
(45 
  per bin) are dark-green and the lowest densities (one per bin) are blue.
  Objects with H$\alpha$ excesses are drawn as magenta circles.}
\label{radec}
 \end{figure}

Figure\,\ref{radec} shows the spatial distribution of VPHAS+ objects in the NW-VMR region.
Even though not the whole region is  covered by VPHAS+ observations, the three \hii regions we study are completely
covered, except for the north part of RCW\,27.  The 329 objects selected as H$\alpha$ emission line stars
are overplotted and clearly show a peculiar pattern with a tight concentration in RCW\,32, a round but quite sparse
distribution in RCW\,27 and an elongated concentration in the RCW\,33 region. This distribution is very similar
to that found for the H$\alpha$-emitting stars by \citet{pett94}.

\begin{figure}
 \centering
 \includegraphics[width=9.5cm]{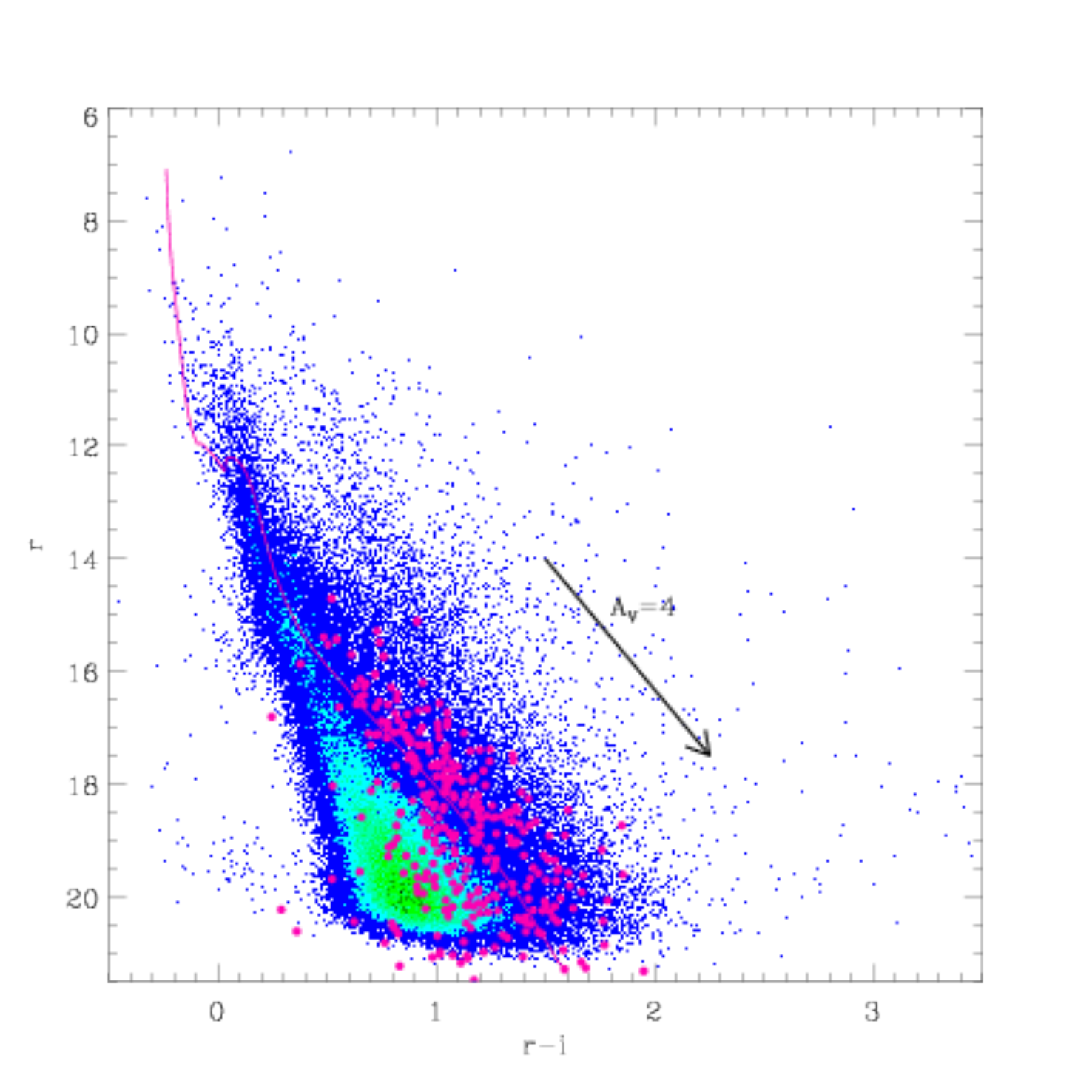}
\caption{SDSS r vs. r-i diagram plotted as a two-dimensional stellar-density histogram 
using a binning of 0.01$\times$0.01 and a 7 level colour-coded scale such that high source densities 
(70 
  per bin) are dark-green and the lowest densities (one per bin) are blue.
  The magenta solid line is the 10\,Myr PISA isochrone, assuming a distance of 750\,pc and E(B-V)=0.2.
  Objects with H$\alpha$ excesses are drawn as magenta circles.}
\label{rri}
 \end{figure}

In order to understand the nature of the selected objects, we plotted them in the r vs. r-i diagram  shown in  
Fig.\,\ref{rri}. Most of the selected objects are on the red side of the diagram and are consistent with a low mass population
of PMS stars, with ages younger than 10\,Myr, as derived from the comparison with the PISA 10\,Myr 
isochrone \citep{togn11,rand17}, assuming a distance
of 750\,pc and  reddening E(B-V)=0.2 (see Section\,\ref{kinematic}). These results
strongly suggest the presence of three  young clusters with different spatial distributions 
associated to the \hii regions RCW\,33, RCW\,32 and RCW\,27, respectively.

\section{Candidate disk-bearing young stars from NIR excesses \label{irexcesses}}
   \begin{figure}[!h]
 \centering
 \includegraphics[width=9.5cm]{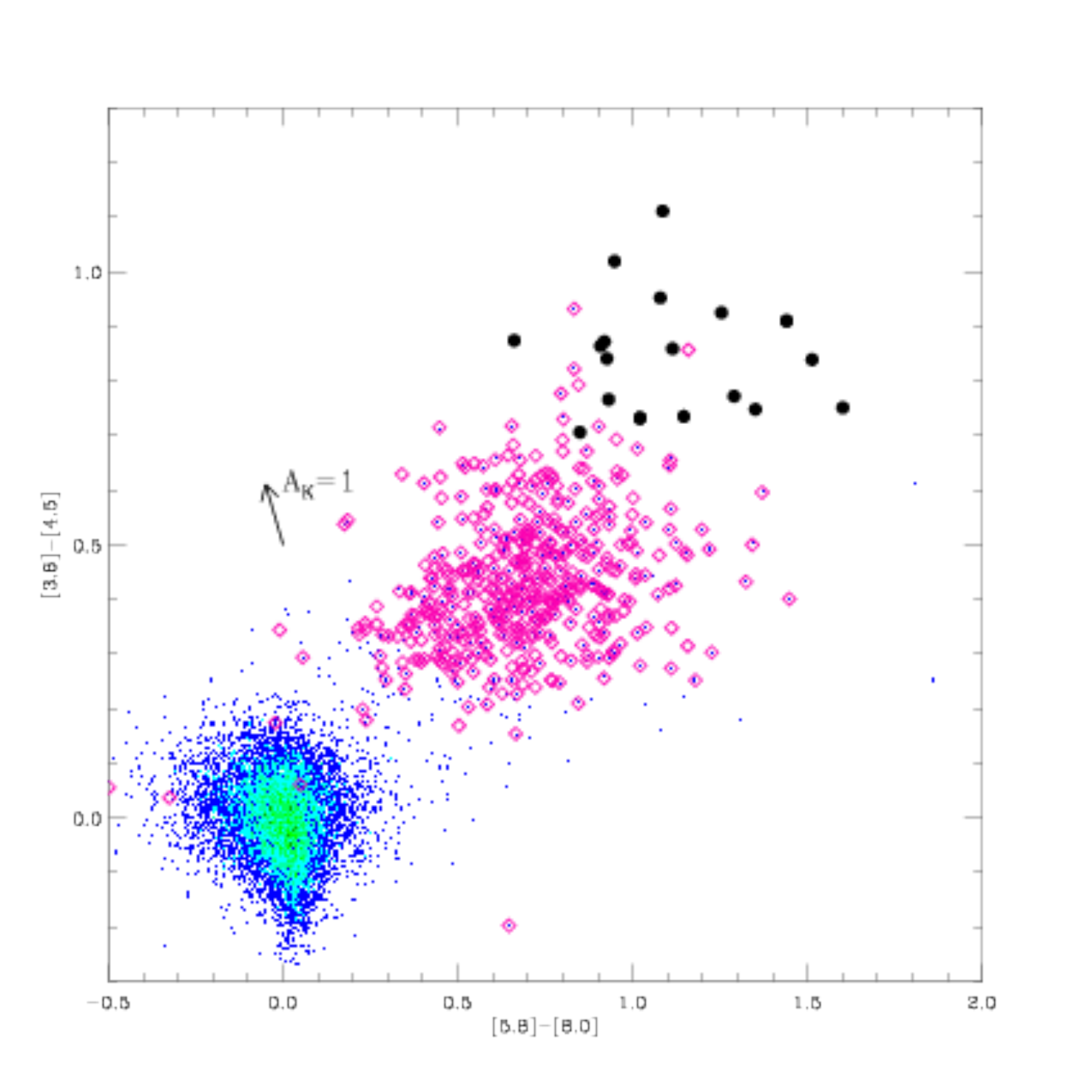}
\caption{[3.6]-[4.5] vs. [5.8]-[8.0] diagram plotted as a two-dimensional stellar-density histogram 
using a binning of 0.01$\times$0.01 and a 7 level colour-coded scale such that high source densities 
(29 
  per bin) are dark-green and the lowest densities (one per bin) are blue.
  Class\,II stars are drawn as magenta diamonds while Class\,I objects are indicated by black points.
  Reddening vector for A$_K$=1 based on the extinction law in the direction of RCW\,49 
  of \citet{inde05} is shown as a filled arrow.}
\label{irexselirac}
 \end{figure}
   \begin{figure}
 \centering
 \includegraphics[width=9.5cm]{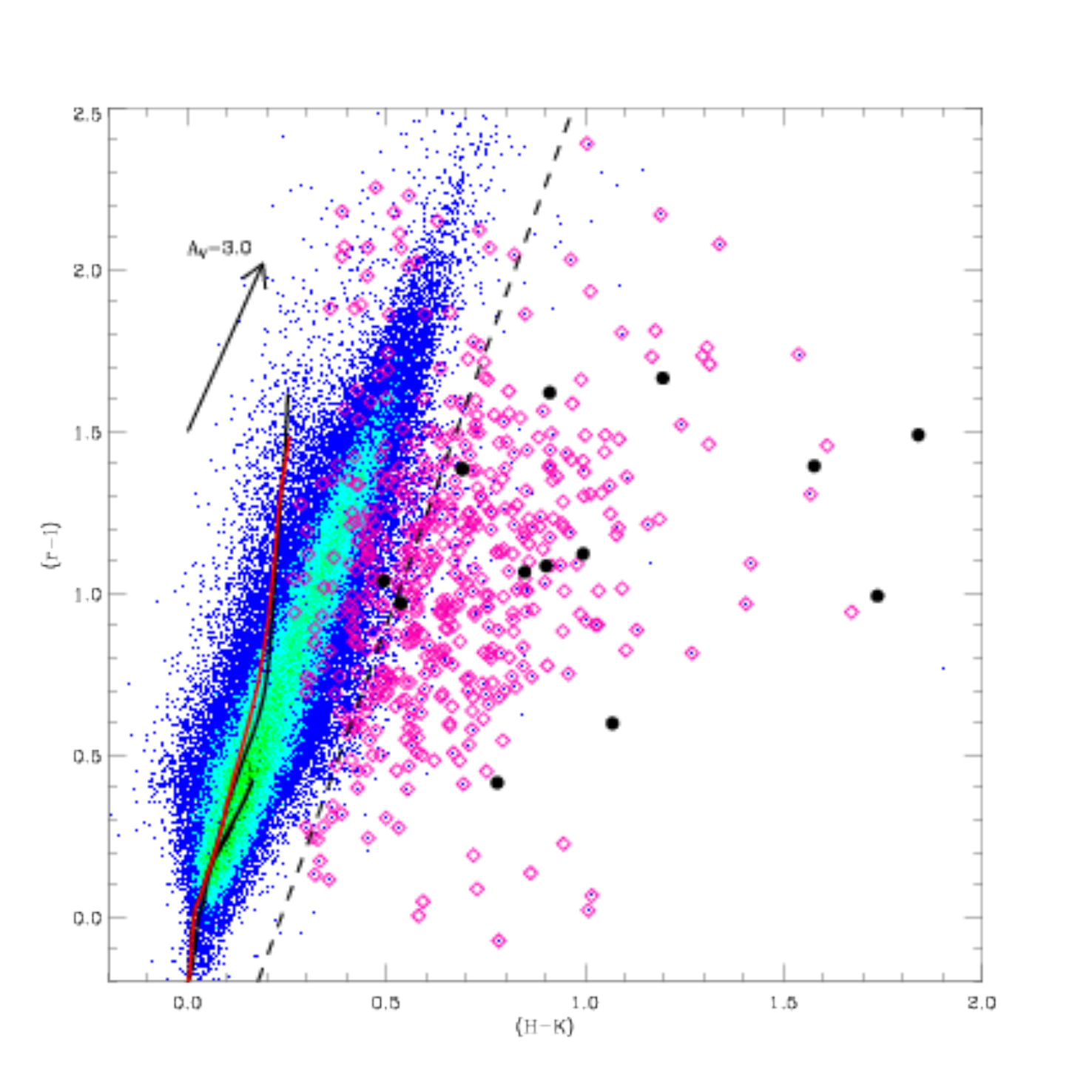}
\caption{r-i vs. H-K diagram plotted as a two-dimensional stellar-density histogram 
using a binning of 0.01$\times$0.01 and a 7 level colour-coded scale such that high source densities 
(54 
  per bin) are dark-green and the lowest densities (one per bin) are blue.
  Symbols are as in Fig.\,\ref{irexselirac}.  The black and red solid lines are the 100 and 10\,Myr 
  PISA isochrones.
  The reddening vector is also shown.}
\label{irexselrihk}
 \end{figure}
  \begin{figure}[!h]
 \centering
 \includegraphics[width=9.5cm]{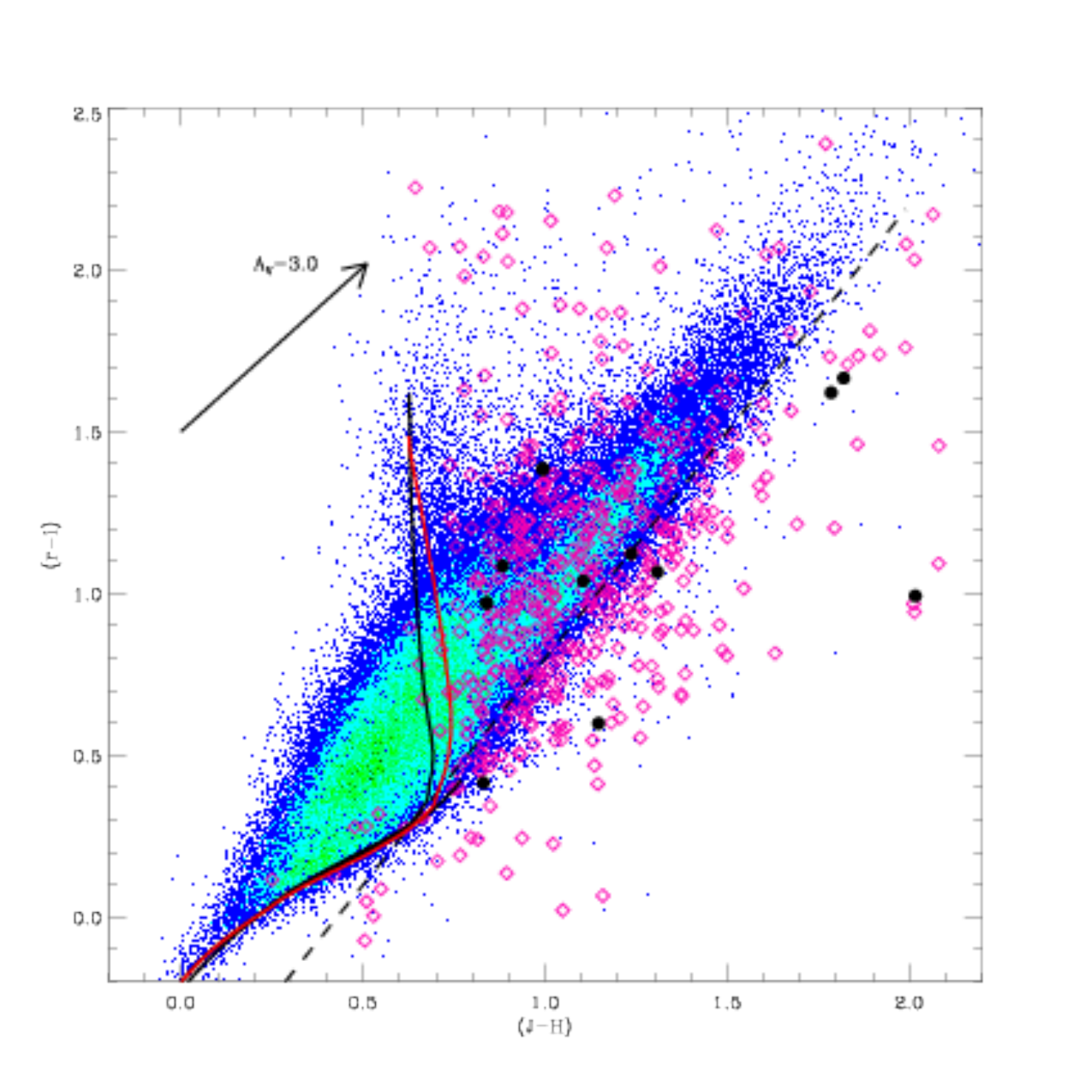}
\caption{r-i vs. J-H diagram plotted as a two-dimensional stellar-density histogram 
using a binning of 0.01$\times$0.01 and a 7 level colour-coded scale such that high source densities 
(39 
  per bin) are dark-green and the lowest densities (one per bin) are blue.
Symbols are as in Fig.\,\ref{irexselirac}.}
\label{irexselrijh}
 \end{figure}
  \begin{figure}[!h]
 \centering
 \includegraphics[width=9.5cm]{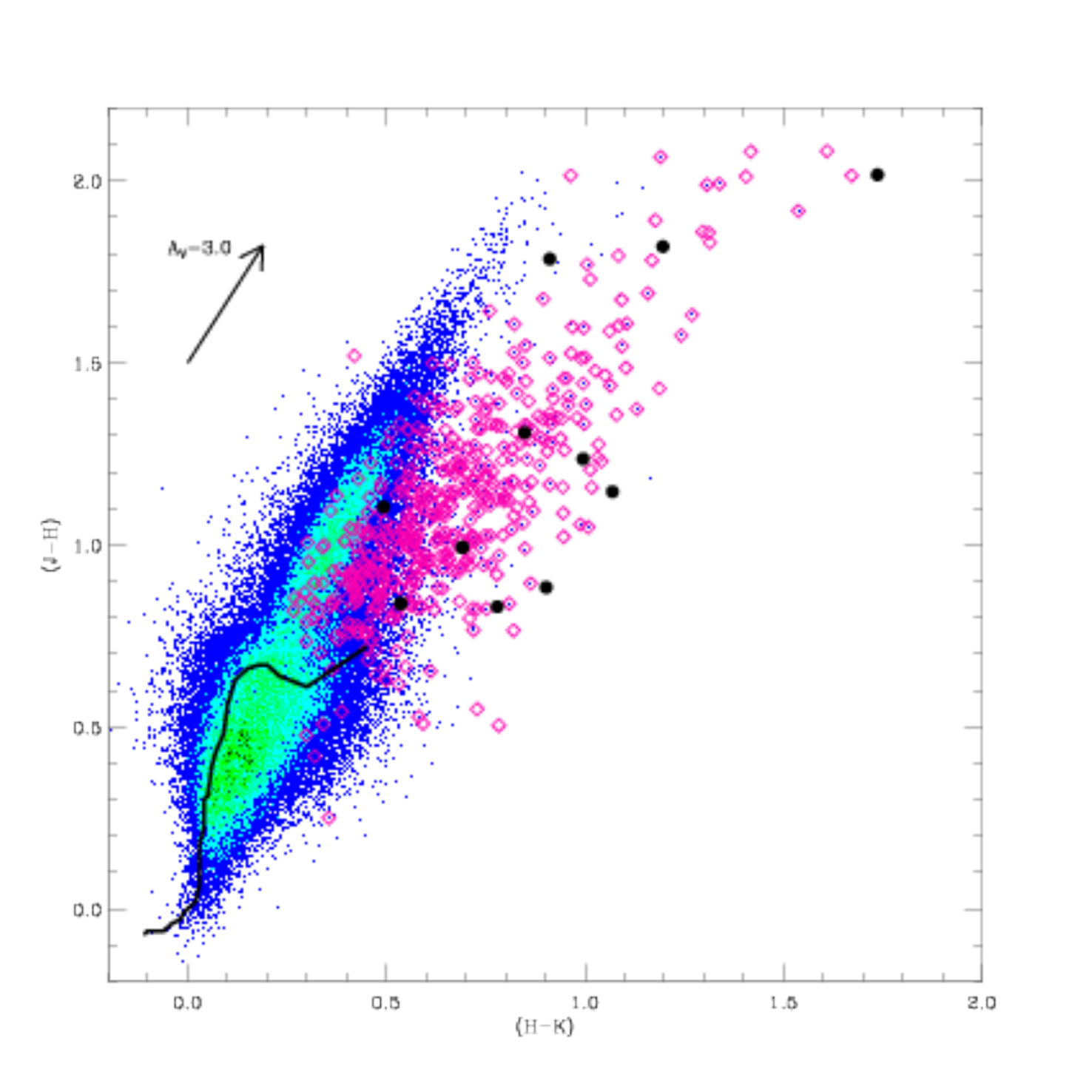}
\caption{J-H vs. H-K diagram plotted as a two-dimensional stellar-density histogram 
using a binning of 0.01$\times$0.01 and a 7 level colour-coded scale such that high source densities 
(56 
  per bin) are dark-green and the lowest densities (one per bin) are blue.
  Symbols are as in Fig.\,\ref{irexselirac}.
  The black solid line is the MS locus \citep{keny95}.}
\label{irexseljhhk}
 \end{figure}
  \begin{figure}[!h]
 \centering
 \includegraphics[width=9.5cm]{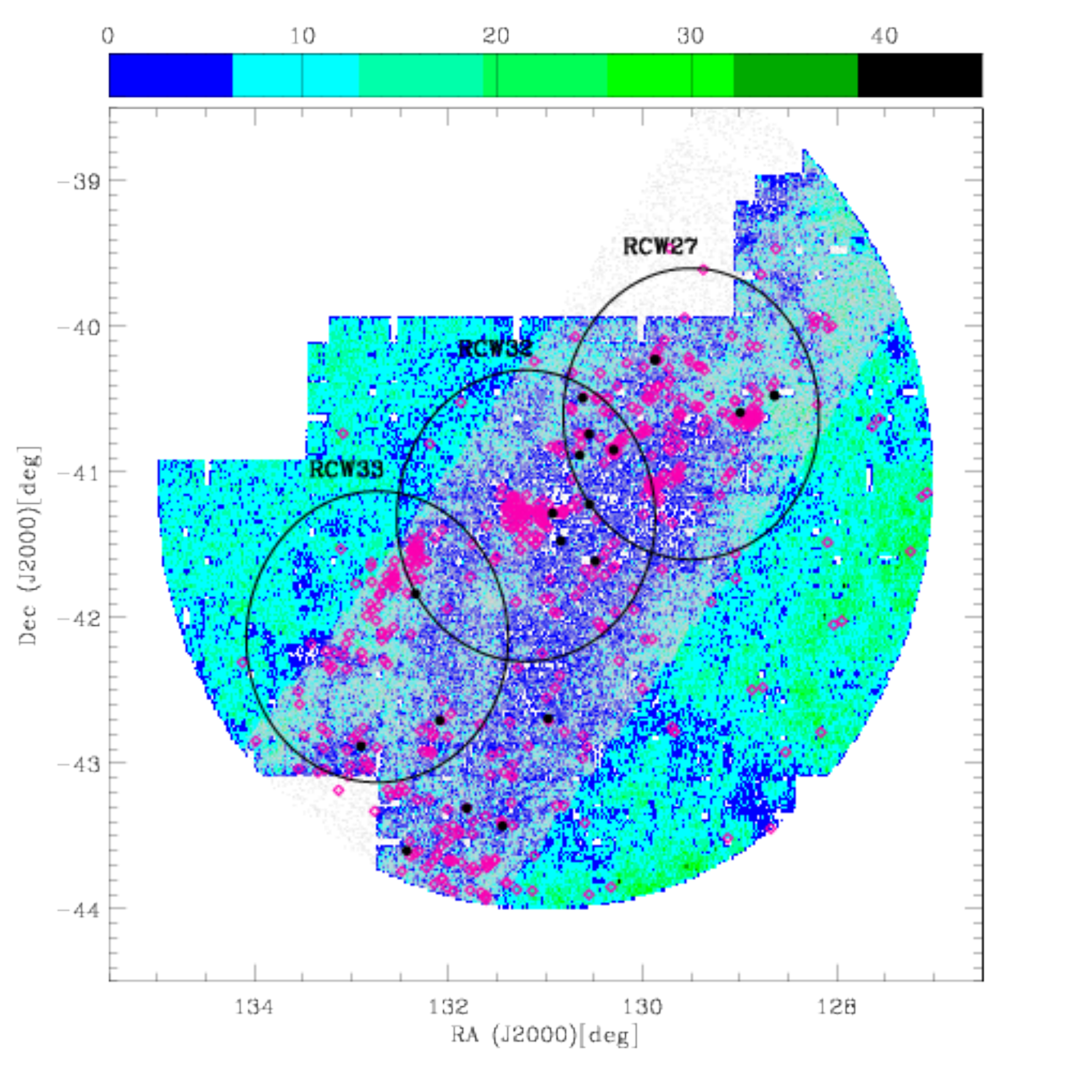}
\caption{Spatial distribution as in Fig.\,\ref{radec} where all the Spitzer/IRAC sources are overplotted with grey symbols.
YOSs with IR excesses are indicated as in Fig.\,\ref{irexselirac}.  The three circular boundaries delimit the regions of
the three clusters.}
\label{radecirex}
 \end{figure}

  \begin{figure}
 \centering
 \includegraphics[width=8.2cm]{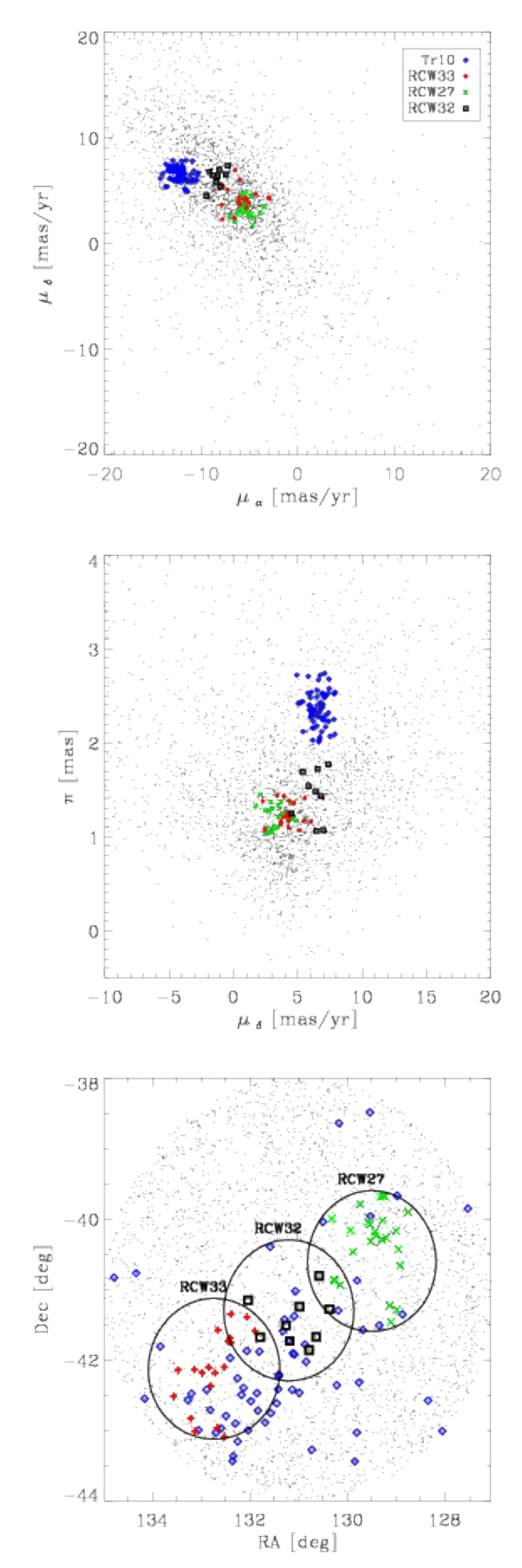}
\caption{$\mu_\alpha$ vs. $\mu_\delta$ (upper panel), $\pi$ vs. $\mu_\delta$ (middle panel) and
spatial distribution (lower panel) of the TGAS catalog in the region of the VMR studied here.
Blue diamonds are associated with the Trumpler 10 cluster,
red plus are associated with YSOs in the RCW\,33 region, 
green X are associated with YSOs in the RCW\,27 region and
black squares are associated with YSOs in the RCW\,32 region. }
\label{kinematicall}
 \end{figure}
 
The presence of NIR excesses with respect to photospheric stellar colors   is a very efficient tool 
to select YSOs. Several optical/NIR colors can be used to select objects with
evident NIR excesses due to the presence of a  circumstellar dusty disk in YSOs.
Spitzer/IRAC CCDs are a powerful method to distinguish YSOs with disk from old stars,
since IR excesses in these colors do not follow the reddening direction and can therefore 
be easily separated from reddened objects. 
The [3.6]-[4.5] vs. [5.8]-[8.0] diagram in Fig.\,\ref{irexselirac} includes all the objects with  error 
 smaller than 0.2 mag and with S/N$>$10 in all four bands, as done in \citet{zaso09} for data from the same survey. 
By using the conditions given in \citet{gute09},
that involves all  IRAC/Spitzer magnitudes,
 we selected   461 candidate Class\,II, i.e. disk-bearing objects  and 20 Class\,I YSOs, i.e. 
YSOs still surrounded by infalling material from which they form.

We also looked for possible contaminants (AGN or PAH galaxies) by following the definitions given in \citet{gute09}.
We found that our sample of Class\,II YSOs includes 15 sources that are likely PAH contaminants 
and we removed them from our sample. 

Warm disks can show IR excesses  in the K and H bands but also in the J band
and, in fact, the J-H vs. H-K diagram is usually considered to select Class\,II stars \citep{meye97,ciez05}.
 However, in this diagram it is very difficult to distinguish reddened objects from NIR excesses, since the
 region consistent with    Class\,II stars overlaps with that of reddened objects, and therefore only objects with 
very strong NIR excesses can be selected from this diagram. 
On the contrary, objects with  excesses in the H or K bands can be very well discriminated 
 from reddened objects if the diagram involves 
an optical color, dominated by photospheric emission even in Class\,II stars.
For example, as discussed in \citet{dami17},
 in the r-i vs. J-H and/or r-i vs. H-K diagrams, objects with excesses in the H-K or J-H colors can be confidently
 found on the red side of the main locus of normal reddened stars. 
 
 Figures\,\ref{irexselrihk}, \ref{irexselrijh} and \ref{irexseljhhk} show the r-i vs. H-K, the r-i vs. J-H and the J-H vs. H-K 
diagrams of all  objects with error in  r-i, H-K and  J-H smaller than 0.1.
It is evident that the bulk of the objects,  mostly including 
giants and MS  stars, follows the reddening vector. 
In this work we adopted the relations of the extinction  curve with R$_V$=3.1  given in Table\,\ref{extinctiontab}.

\begin{table}
\caption{Relations for a G2V star using the \citet{card89} and \citet{odon94} extinction curve with R$_V$=3.1 
in the SDSS ugriz and 2MASS photometric systems \citep{mari17}.}             
\label{extinctiontab}      
\centering                          
\begin{tabular}{c c c c c c c c c}        
\hline\hline             
$\frac{A(u)}{A(V)}$&$\frac{A(g)}{A(V)}$ &$\frac{A(r)}{A(V)}$ &$\frac{A(i)}{A(V)}$ &$\frac{A(z)}{A(V)}$&$\frac{A(J)}{A(V)}$ &
$\frac{A(H)}{A(V)}$ &$\frac{A(Ks)}{A(V)}$ \\    
\hline   
  1.569 &   1.206 &   0.871 &   0.683 &   0.492 &   0.294 &   0.181 &   0.118   \\                  

\hline                                   
\end{tabular}
\end{table}

The comparison of the data with the 10 and 100\,Myr isochrones
suggests that the tail of objects in Fig.\ref{irexselrijh} with J-H around 0.6 and r-i$\gtrsim$1
is the locus of M-type stars, while
 objects on  the right of 
 the dashed lines in the first two diagrams can be considered candidate YSOs. There are in total 221
YSOs showing excesses both in the J-H and H-K colors. Among them, 
111 were  selected as YSOs with disk also  using the IRAC data,
 while the remaining 110 lie outside the region covered by the IRAC/Spitzer
observations. This finding  is fully consistent with the observation coverage. 

To include also the  objects that lack detections in the less sensitive 
 5.8 and 8.0\,$\mu$m bands, or with small excess in the J band, we selected  296 additional
 YSOs by taking objects with 
 excess in the H-K color and that fulfill at least one of the previous three conditions involving the IRAC colors
 [3.6]-[4.5], [3.6]-[5.8] and [4.5]-[8.0].  Finally, to include possible YSOs with a small excess in the K band,
 we selected another 125 YSOs, with  excess in  J-H and in at least one of the IRAC colors.
 At the end, we have a total of 559 Class\,II 
  YSOs, included in at least one of the samples selected as described above,
  plus 20 Class\,I YSOs that are  shown in
  Fig.\,\ref{radecirex},
  where the area covered by IRAC/Spitzer observations is also shown. 
The spatial distribution of the  Class\,II and I stars is very similar to 
that found in the previous section for CTTS  and, again, confirms the presence of three 
embedded young clusters in NW-VMR.  The circular boundaries shown in the figure 
 trace the locations of the dusty shells/bubbles that surround the three \hii regions, 
as can be seen, on the WISE image of the region (see Fig.\,\ref{magebinimg}).   
    \begin{figure*}[!h]
 \centering
 \includegraphics[width=16cm]{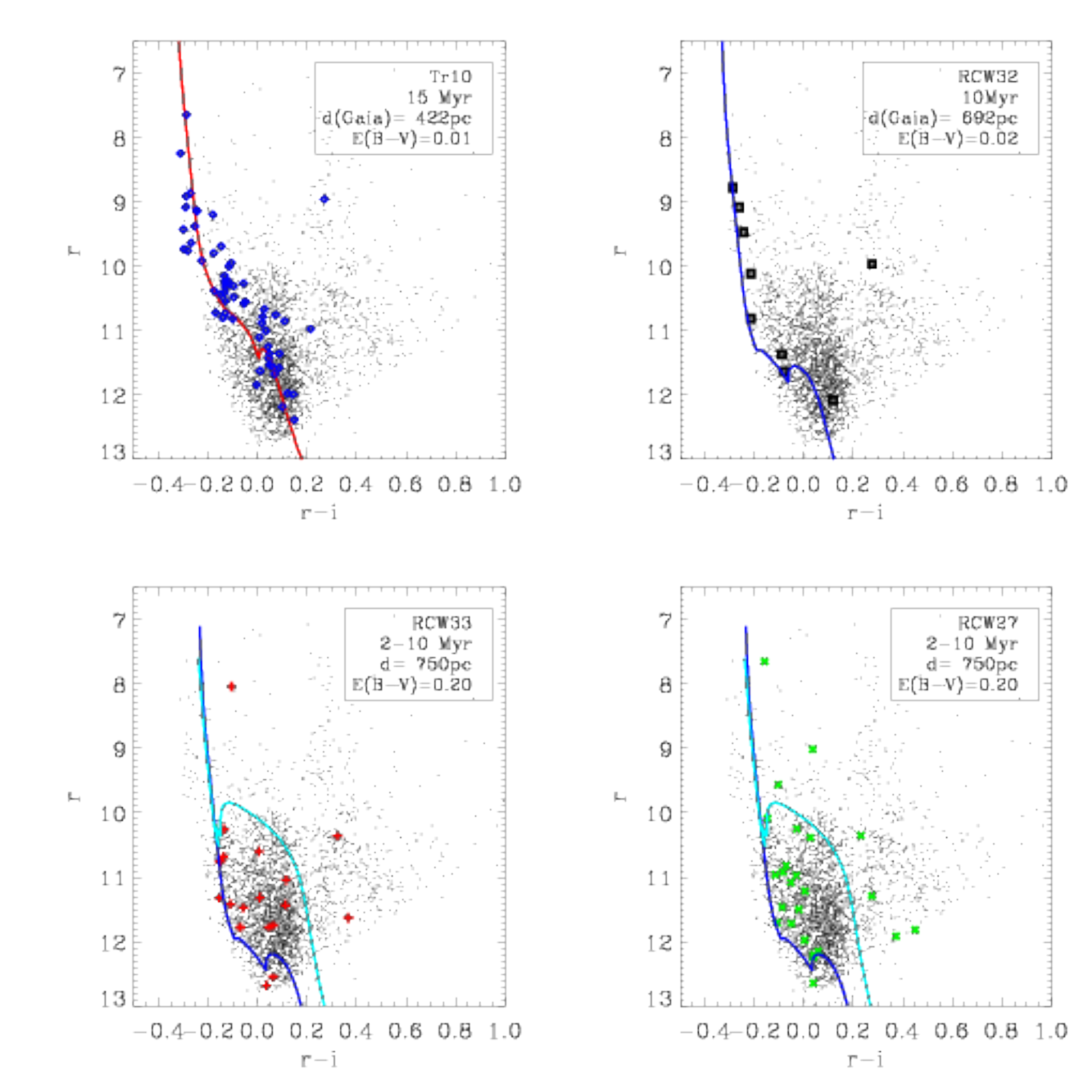}
\caption{r vs. r-i diagrams of the cluster Trumpler 10 and of the three regions centered on 
RCW\,33, RCW\,27 and RCW\,32. Coloured symbols are as in Fig.\,\ref{kinematicall} and 
 indicate the candidate members for each region, selected kinematically
by using Gaia data.}
\label{kinematiccmd}
 \end{figure*}

\section{Kinematic populations\label{kinematic}}
\subsection{The Trumpler 10 association}

Even if limited to very bright stars, Gaia TGAS data give us the opportunity to study the 
kinematics across large fields. In order to understand if the three young populations
belong to the same molecular cloud  and  
share  a similar motion, we compared their PMs and parallaxes.
As already known in the literature, the VMR  is behind the intermediate age OB association
Trumpler\,10   (RA=131\fdg872, Dec=-42\fdg40) of
  which 23 members, 22 B-type and one A0V star are known \citep{de-z99,khar13}. 
 
The distance of Trumpler 10 has been estimated by several authors: 
 417 pc  \citep{khar13}, 424 pc \citep{dias02}, 366$\pm$23 pc \citep{de-z99}.
Located at a galactocentric distance of  
R$_{GC}$ = 8.1 kpc, it suffers from a relatively low reddening, E(B-V) = 0.029  \citep{khar13}. 
The most recent values for the cluster age are 
 24 Myr \citep{khar13}, $\sim$ 35 Myr  \citep{dias02} and  15 Myr  \citep{de-z99}.

Since our field includes this association, we first studied the PM and parallax distributions around its
center, as  shown in Fig.\,\ref{kinematicall}. To detect a population sharing the same motion and with a common parallax,
we adopted the following procedure: we computed the maximum of a bidimensional array of PMs in RA and 
Dec, assuming a PM bin of 1\,mas/yr both in $\mu_\alpha$ and $\mu_\delta$ in the range $-20<\mu_\alpha/$mas yr$^{-1}<20$  
and $-20<\mu_\delta/$mas yr$^{-1}<20$  and  in parallax steps of $\Delta \pi$=1\,mas. This
large $\pi$ step is chosen to account for the large errors on the parallax values of the 
TGAS GAIA data. 
To avoid a binning dependence, we repeated the computation by shifting by 0.1\,mas
the parallax bin  while  exploring the parallax range $0<\pi/$mas$<4$,
(corresponding to distances larger than 250\,pc) where most of the objects are distributed. 
The maximum value of each array represents the tridimensional condition to detect clusters of objects at the
same distance sharing the same motion. For each parallax bin, 
we considered the   $\mu_\alpha$ and $\mu_\delta$
corresponding to the maximum of the array and therefore we selected all the objects with 
$\sqrt{(\mu_\alpha)^2+(\mu_\delta)^2}<$2\,mas.
If the number of objects within this 3-d region is larger 
than 10, then
the corresponding $\mu_\alpha$, $\mu_\delta$ and $\pi$ are associated with a population\footnote{the adopted
 threshold corresponds to about 3$\sigma$ of the distribution of objects where no hint of clustering
is found}.
As expected, we obtained similar values for close parallax ranges    and therefore we considered
as final  parameters for a population, the median values associated 
with the selected maxima.  

To detect the population of Trumpler 10, we considered the whole TGAS catalog in the entire region
we are studying,  because this association is known to be very dispersed on the sky.
In this region, our procedure enabled us to detect a main  population,
at  $\pi=2.37$ mas, corresponding to 422$\pm$32\,pc (1\,$\sigma$ uncertainty). 
We conclude that this latter population  
is the one associated with Trumpler 10 and includes 63 objects. 

Figure\,\ref{kinematicall} shows also
the spatial distribution of the TGAS stars found in this region, where the population of Trumpler 10 
(blue symbols) is well distinct from the others (see below) for its kinematic parameters, while its spatial
distribution is very dispersed throughout the region, confirming the  \citet{de-z99} results.
The color-magnitude diagram (CMD) of these objects is shown in Fig.\,\ref{kinematiccmd} and compared with the
PISA isochrone with the parameters indicated in the Figure. The sequence shows
a spread around the isochrone. A  thorough assessment of the nature of this spread, 
which would require detailed investigation of the  entire population of the association,
 is beyond the aim of this work.

\subsection{The cluster Cr\,197 and RCW 32} 
The analysis of PMs and parallaxes in the region of RCW\,32, corresponding  to the cluster Cr\,197, has been performed
as done for Trumpler 10. This last  association   is spread across a large region  overlapping
that of Cr\,197.
For the kinematic analysis of Cr\,197, we used only the TGAS data
  within a radius of 45\arcmin\ from its center.
  This radius has been chosen to avoid
 the two neighbouring  regions RCW\,33 and RCW\,27, that show a
  more dispersed spatial distribution.
 In addition, we discarded  objects with $\mu_\alpha<-10$\,mas/yr
 and $\pi>2$\,mas, since they are associated with Trumpler\,10. Our procedure enabled us to detect two groups of objects in the
  PM plane, lying in the same parallax range. 
The two populations do not show a peculiar spatial distribution, being both quite sparsely distributed, but their 
distribution in the CMD suggests that the group of objects corresponding to 
($\mu_\alpha$, $\mu_\delta$)=(-8.3, 5.8) mas/yr in the parallax range $1.0<\pi<$2.0
(indicated with black squares in  Fig.\,\ref{kinematicall})
can be associated with the cluster Cr\,197. The mean parallax of these objects is 1.44 mas, which corresponds to 
a distance of 692$\pm$ 128\,pc.  
 Indeed, the  9 selected objects (black squares in Fig.\,\ref{kinematicall} and \ref{kinematiccmd}) follow quite well the 10\,Myr PISA isochrone,
  at  a distance of  692$\pm$80\,pc ($\pi=1.45$), assuming E(B-V)=0.02, as shown in Fig.\,\ref{kinematiccmd}.
 Our analysis therefore suggests a cluster distance significantly smaller
than that found by \citet{bona10}, namely 1050$\pm$200\,pc,
 based on near-infrared photometry and a statistical field star decontamination. Our value is, instead, consistent 
 with the distance of $\simeq$700\,pc derived by \citet{geor73} and \citet{cram74}.
  In addition, the r vs. r-i diagram suggests that the cluster
 is less  reddened than the value reported in the literature, equal to E(B-V)=0.34 \citep{bona10}.

\subsection{The cluster Vela T2 and RCW 33}
As done in the previous sections, we performed the PM and parallax analysis also in the RCW\,33 region.
Even in this case, there is a strong contamination from the Trumpler 10 members
and therefore we discarded  objects  with $\mu_\alpha<-10$\,mas/yr  and $\pi>2$\,mas. 
After that, the diagram ($\mu_\alpha, \mu_\delta$)  shows a group of objects
with a quite elongated distribution in the  $\mu_\alpha$ axis. This  includes two groups that are  
distinguishable in the parallax diagram,  in the range $0.5<\pi$/mas$<1.0$ and  $1.0<\pi$/mas$<1.5$.
The second subgroup, corresponding to a distance of about 800\,pc,  is centered   
to ($\mu_\alpha$, $\mu_\delta$)=(-5.57, 4.10)\,mas/yr
(red symbols in Fig.\,\ref{kinematicall} and \ref{kinematiccmd}). This population includes 19 objects with an elongated 
spatial distribution 
similar to the YSOs selected above.
   In the CMD, these objects lie in a region 
compatible with both MS and PMS stars
(Fig.\,\ref{kinematiccmd}). We associate this group of stars to the cluster Vela T2.

\subsection{The cluster Vela T1 and  RCW 27} 
The PM analysis of the objects in  RCW 27 shows a well distinct clump in 
the  ($\mu_\alpha$, $\mu_\delta$) diagram for $\pi>1.0$\,mas\,
at ($\mu_\alpha$, $\mu_\delta$)=(-5.48, 3.39)\,mas/yr.
The  population, corresponding  to  RCW\,27 has been selected by taking objects
with PMs within 2\,mas/yr from this peak and 
  with $1.0<\pi<1.5$\,mas.
This  population includes 25 objects
that mainly lie in the north-western part  of RCW\,27 (green symbols in Fig.\,\ref{kinematicall} and \ref{kinematiccmd}).
 The CMD
of the selected objects is consistent with a PMS population located at about  750 pc, with age less than 10\,Myr. 

\section{Summary of "classical" YSO selecion \label{classicalmembership}}
We summarise here the properties of candidate YSOs found in the NW-VMR including  all the objects
with H$\alpha$ or NIR excesses, selected as described in Sect.\,\ref{halphaexcesses} 
 and \ref{irexcesses},
 with a X-ray detection, or with PMs and parallaxes consistent with the populations in RCW\,33, RCW\,32 or RCW\,27,
selected as described in Sect.\,\ref{kinematic}.  

We selected a total of  907 candidate YSOs in  the whole area. Those included in the circular 
regions   with radius 1\,deg in
RCW\,33, RCW\,32 and RCW\,27, are 210, 177 and 288, respectively.
Note that the YSOs falling in the overlapping areas were associated to the external ones,
 i.e. RCW\,33 and RCW\,27. 
We will refer to the  907  objects as members selected with {\it classical} methods.
Optical/NIR photometry and the membership information of  the selected YSOs are given in Table\,\ref{classmemtab}.
\begin{table*}
\caption{Optical/NIR photometry and membership criteria of YSOs selected with {\it classical}  methods.
M1, M2, M3, M4 and M5 stand for membership from H$\alpha$, u, IR excesses, kinematics and X-ray detection, respectively;
 1 indicates member while 0 indicates non-member.
 Full table available in electronic format only.
 \label{classmemtab}}
\centering
\begin{tabular} {c c c c c c c c c c c c c c c c}  
\hline\hline
CNAME & RA  &   Dec & u & g & r & i & H$\alpha$ & J & H & K & M1 & M2 & M3  & M4  & M5\\
      & (J2000) & (J2000)    &  &  &  & & &  &  &  & &  &   &  & \\
\hline
08442605-4105159 &    131.10853 &   -41.08775 & ...  & ...  &   11.43 &   11.22 & ...  &   10.19 &    9.80 &    9.69 & 0 & 0 & 0 & 0 & 1\\
08435467-4114538 &    130.97779 &   -41.24827 & ...  & ...  &    9.46 &    9.71 & ...  &    9.33 &    9.35 &    9.30 & 0 & 0 & 0 & 1 & 0\\
08441101-4116189 &    131.04588 &   -41.27192 & ...  & ...  &   11.65 &   11.20 & ...  &   10.22 & ...  & ...  & 0 & 0 & 0 & 0 & 1\\
08443774-4115371 &    131.15724 &   -41.26032 & ...  & ...  &   14.75 &   14.11 & ...  & ... & ...  & ...  & 0 & 0 & 0 & 0 & 1\\
08444029-4116378 &    131.16788 &   -41.27716 & ...  & ...  &    7.24 &    7.23 & ...  &    6.39 &    6.25 &    6.24 & 0 & 0 & 0 & 0 & 1\\
08451790-4112462 &    131.32456 &   -41.21283 & ...  & ...  &   14.90 &   14.27 & ...  & ... & ...  & ...  & 0 & 0 & 0 & 0 & 1\\
08444546-4117322 &    131.18942 &   -41.29228 & ...  & ...  &   14.57 &   13.94 & ...  & ... & ...  & ...  & 0 & 0 & 0 & 0 & 1\\
08451500-4117349 &    131.31251 &   -41.29301 & ...  & ...  &   12.19 &   12.05 & ...  &   11.20 &   10.93 &   10.87 & 0 & 0 & 0 & 0 & 1\\
08450273-4123123 &    131.26138 &   -41.38674 & ...  & ...  &   15.67 &   15.13 & ...  & ... & ...  & ...  & 0 & 0 & 0 & 0 & 1\\
08445256-4124287 &    131.21902 &   -41.40797 & ...  & ...  &    8.78 &    8.71 & ...  &    7.83 &    7.55 &    7.48 & 0 & 0 & 0 & 0 & 1\\
08411684-4050026 &    130.32015 &   -40.83406 & ...  & ...  &   11.22 &   11.09 & ...  &   10.11 &    9.69 &    9.37 & 0 & 0 & 1 & 0 & 0\\
\hline
\end{tabular}
\end{table*}
 
\begin{figure}
 \centering
 \includegraphics[width=9.5cm]{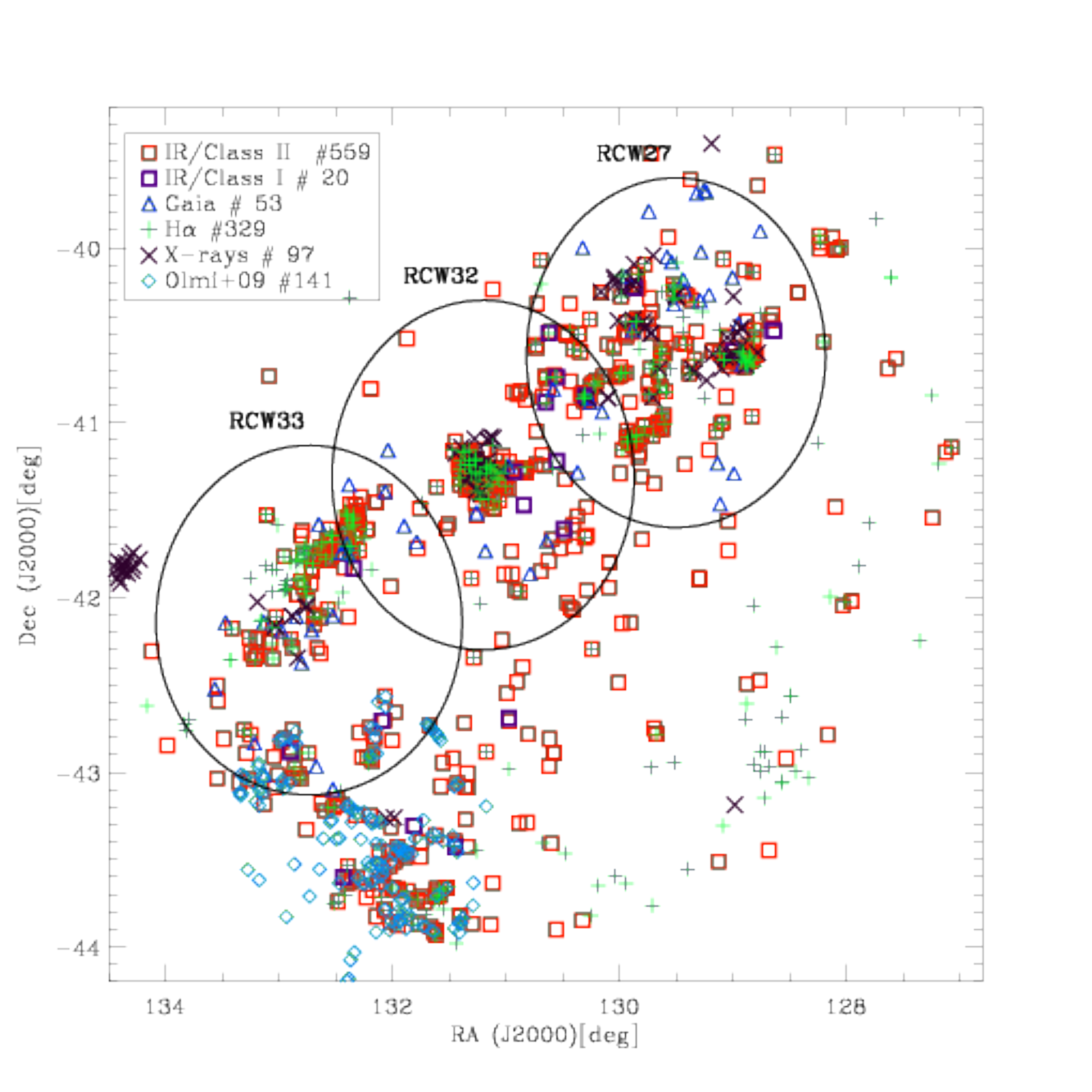}
\caption{Spatial distribution of all candidate YSOs, selected with different membership {\it classical} criteria,
 plotted with different symbols.  Cyan diamonds are the 141 sub-mm cores found with the BLAST survey by \citet{olmi09}.
  The three circles of 1\min in radius delimit the regions around the three clusters.}
\label{radecmem}
 \end{figure}

We note that the list is not complete since the spatial coverage of data used to select them is  
incomplete. The spatial distribution of all the objects selected as YSOs is shown in Figure\,\ref{radecmem}.
 We omitted from Figure\,\ref{radecmem} the 63 kinematic objects candidate members of Trumpler\,10 since they are not
related to the young populations we are investigating in this work.

 As already discussed, many of the selected YSOs define  three young clusters associated to the three \hii regions.
However, there is an
additional stellar sub-cluster located right to the south of RCW 33, corresponding to Vela-D,
 and composed of mainly H$\alpha$ and IR-excess YSO candidates.
This region has been the subject of a Balloon-borne Large-Aperture Submillimeter Telescope (BLAST) survey 
by \citet{olmi09} where 141 pre-stellar and protostellar  sources have been found. The presence of YSOs objects
detected also in the optical and NIR bands and the strong spatial correlation of these
YSOs with the cores found in \citet{olmi09} is a further confirmation  of the ongoing star formation in this region,
characterised by the  lack of massive O-type stars. 

We note the presence of an additional group of X-ray sources located
  to the east of RCW 33. These objects are located in a region devoid of YSOs selected in this work, 
and  on the boundary of  the field and thus we are not able to assert if this group of objects is
 associated to a young sub-cluster.

The three circular areas in the \hii  regions are those with the best coverage but, for example, the 
northern
region of RCW\,27 lacks  VPHAS+ data and the northern regions of RCW\,33 and RCW\,32 lack SPITZER/IRAC data.
For this reason, we will avoid any analysis requiring completeness of the samples.
The YSOs with evidence of accretion from H$\alpha$ and of circumstellar disk from the IR are 136.
The objects with NIR excess, selected also with PMs and parallax, are 3,
while no objects, selected kinematically, show evidence of accretion or X-ray emission.
Most of the candidate YSOs have only one youth indicator and this suggests that the methods are complementary.
 
We find that members selected as YSOs with NIR,  H$\alpha$ excess and  X-rays
 show a  similar spatial distribution in each of the 3 \hii  regions,
even though the X-ray sample is very limited regarding both  spatial coverage and duration of observations.
The members selected using Gaia data, expected to be the more massive ones, 
show a spatial distribution similar to that of the other members in Vela T2 and T1,
while in  Cr\,197, they are  poorly correlated
with the other member positions. This could be due to the lower performance  of Gaia data
in  very crowded regions, as the cluster Cr\,197 in  RCW\,32   or to source extinction. In fact,
the TGAS sample  mainly includes lightly absorbed stars.
Since the region harbors a large number of spatially clustered disk-bearing YSOs, clouds and a nebular
background, a spatially varying source extinction is expected, that could explain the poor correlation
with Gaia members.



In conclusion, we find that the whole region includes different sites of ongoing or recent SF.
We will concentrate our attention on the three areas of the \hii  regions, whose
 properties  will be studied in the 
following sections.

   \begin{figure}
 \centering
 \subfloat{\includegraphics[width=7.5cm]{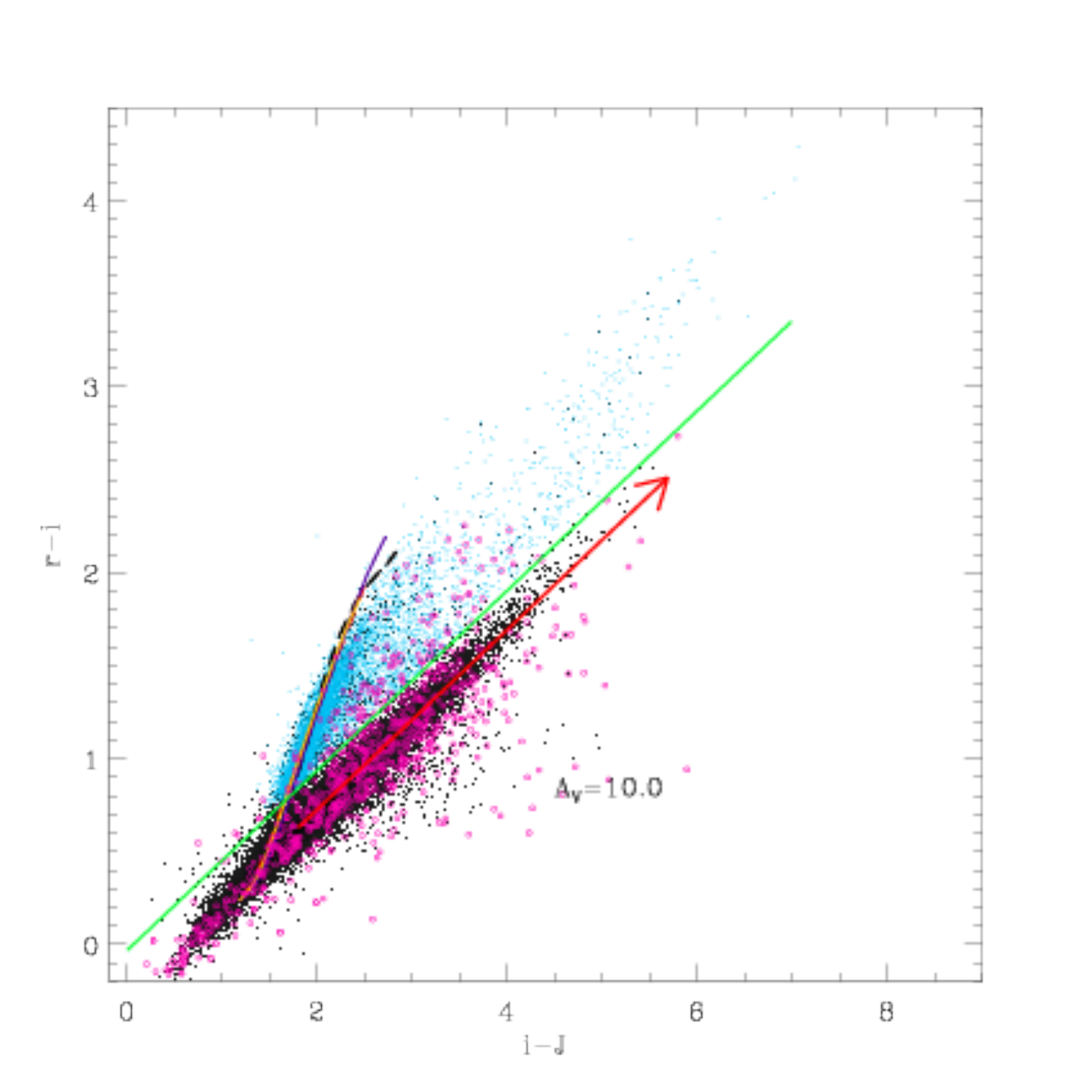}}
 \vspace{0.1cm}
 \subfloat{\includegraphics[width=7.5cm]{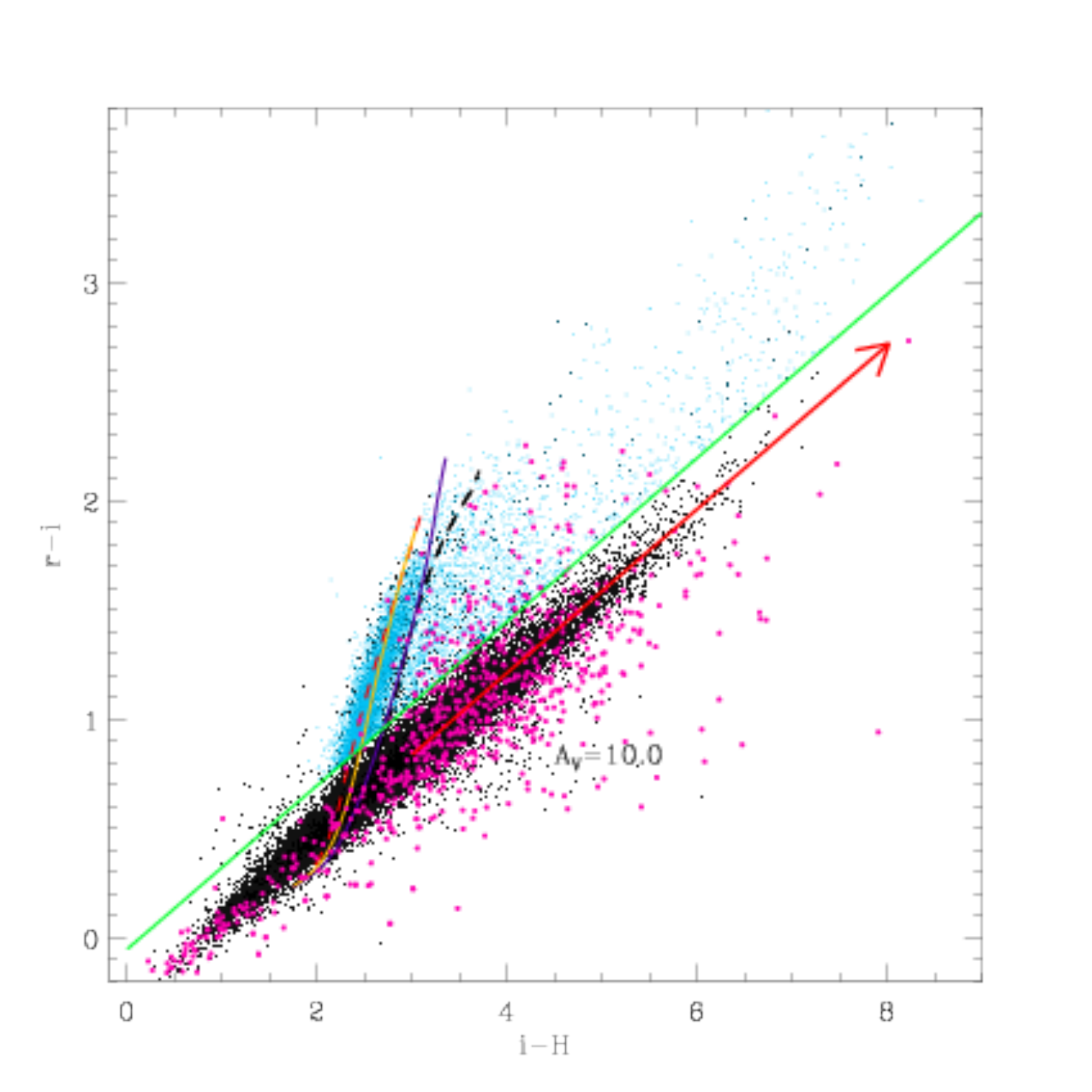}}
  \vspace{0.1cm}
 \subfloat{\includegraphics[width=7.5cm]{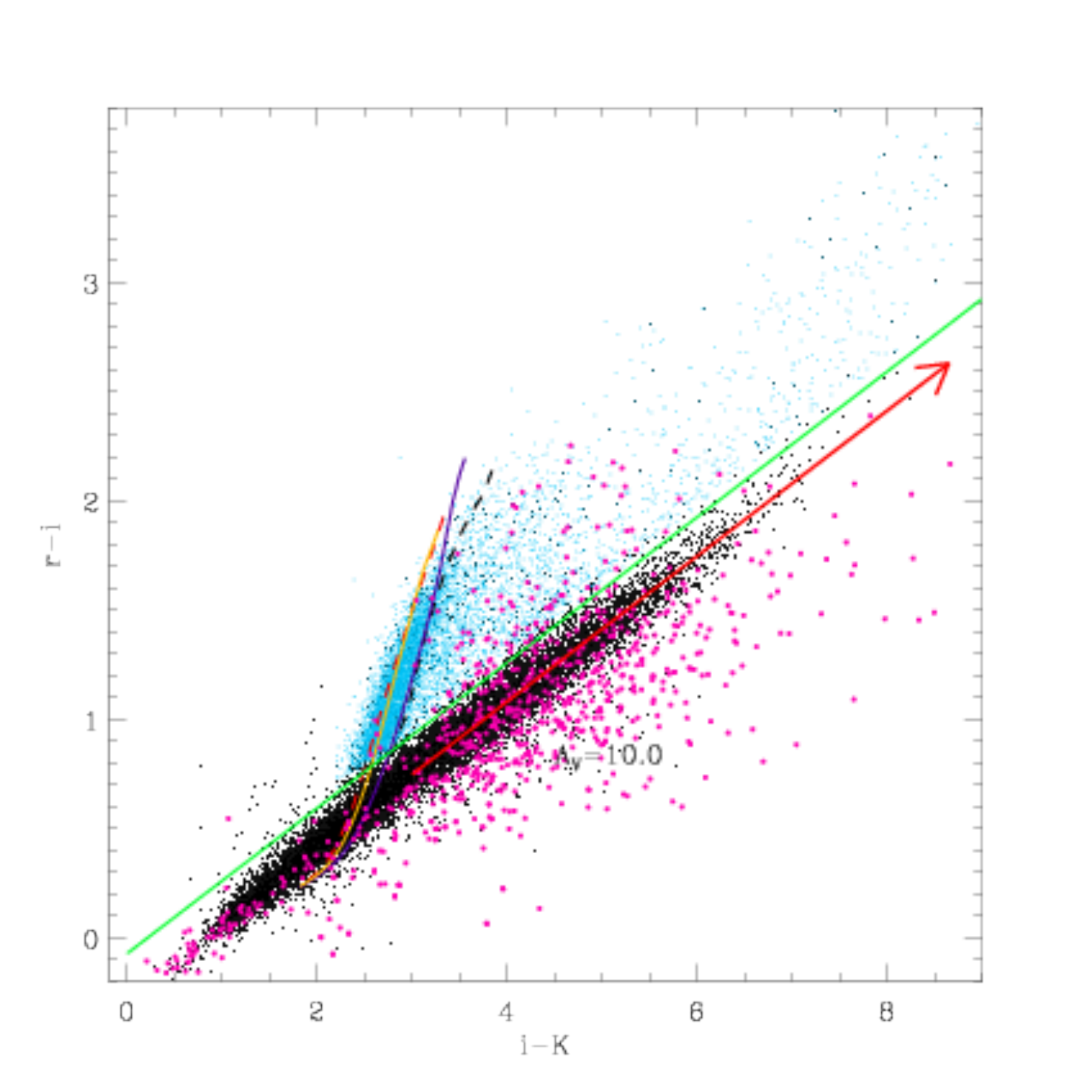}}
\caption{r-i vs. i-J, i-H and i-K  diagrams plotted as a two-dimensional stellar density histogram in a grey scale,
 compared with 
the MS locus (dashed red line), 
the 10\,Myr and 0.5\,Myr isochrones (orange and purple lines)  and
the giant locus (black line) predicted by PISA models. YSOs selected with {\it classical} methods
and objects selected as M-type stars
are indicated with magenta and cyan symbols, respectively.
  The reddening vectors and the limits used for the selection of M-type stars are indicated as  red arrows
  and  green lines, respectively.}
\label{riijmselpisa}
 \end{figure}

\section{Features  of M-type stars \label{mtypestars}}The region studied in this work lies on the galactic plane and therefore includes
 a large number of field stars. Due to the presence of the molecular clouds,
 the  background objects are expected to be very reddened,  while 
 those in front of the nebula are likely affected by small reddening.
 In general, in the optical and NIR CCD and CMD,
 foreground and background field stars cannot be distinguished from YSOs,
since the color distribution of all these objects follows the direction of the reddening vector. 
Nevertheless,   using appropriate combinations of optical and NIR magnitudes, 
the color distribution of M-type  stars does not follow the reddening
vector and this property can be used to select M-type stars,
 as recently discussed in \citet{dami18} and Venuti (2018, in preparation). 

  This is  clearly visible in the r-i vs. i-J, i-H and i-K diagrams shown in
 Fig.\,\ref{riijmselpisa}, where the comparison with the theoretical loci of
 PMS, MS and giant stars  shows that  the region of  M-type stars (r-i$\gtrsim$0.75)
 is clearly distinguishable from that of hotter, reddened stars.
In particular, the bulk of our data in the M-type range is compatible with   stars
affected by very small reddening,
  while all  objects on the red side of the diagram 
are very reddened  stars.

 Following the position of the theoretical loci in the three diagrams, compared to the  bulk of objects 
 that lie along the reddening vectors,
 we select as M-type stars all the objects with r-i larger than the limits indicated by the solid green lines
 in all the three diagrams, and with i-K$>$2.2. 
 The simultaneous use of the three diagrams allows us to select a sample  of M-type stars that is 
 as clean as possible. In the three diagrams,  YSOs selected with {\it classical}
 methods mainly lie in the region of  reddened stars and only
 a small fraction of them fall in the region of M-type stars. As expected, the spread of the selected YSOs    
in the r-i vs. i-K diagram is larger than that in the r-i vs. i-J diagram, since this sample 
includes stars with IR excesses, mainly in the K band. Therefore, it could include 
also M-type stars with disk or accretion, that, due to the presence of IR excesses, fall
outside of the region of M-type stars.
    \begin{figure}
 \centering
 \includegraphics[width=9.5cm]{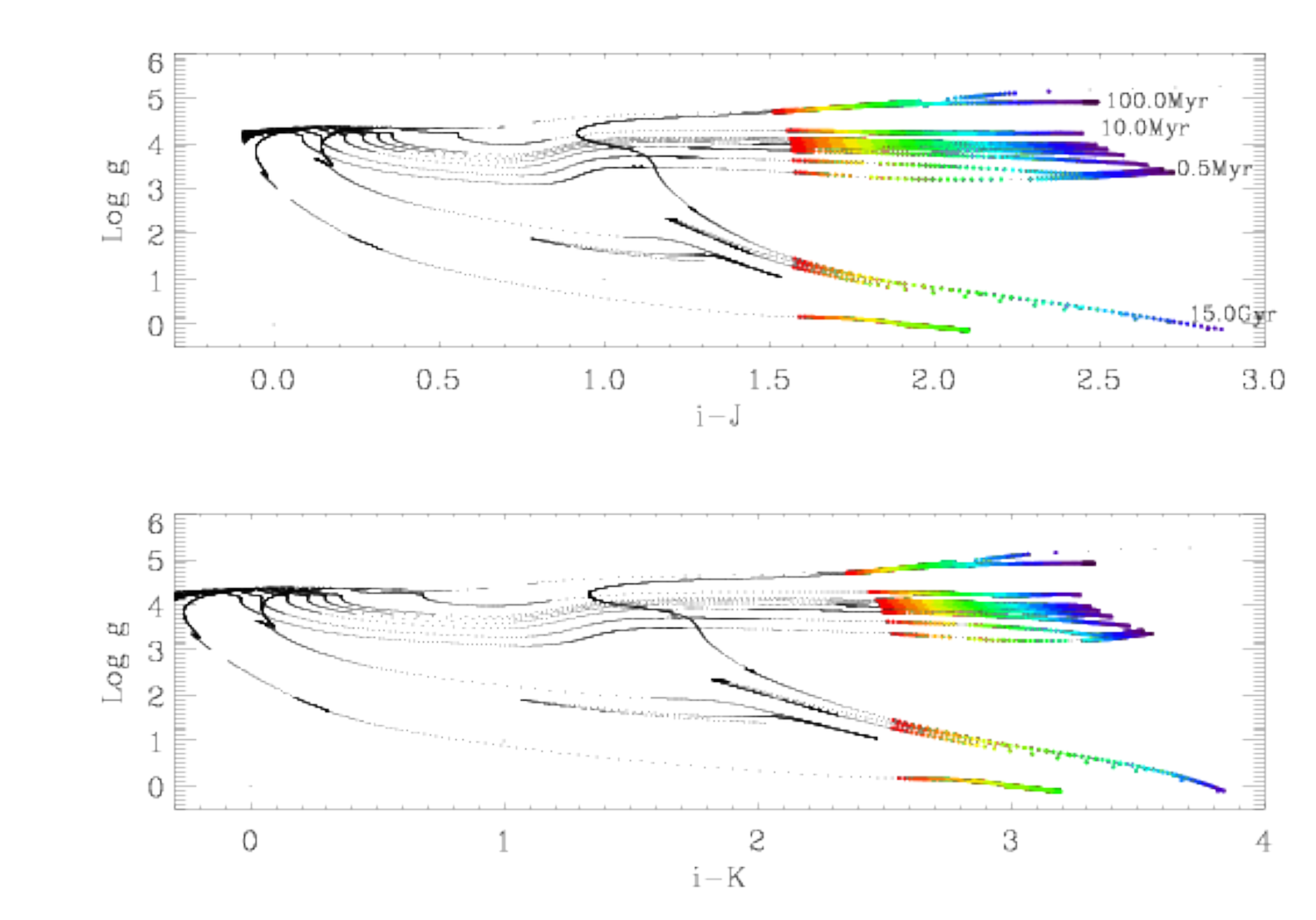}
\caption{Log g values vs.  i-J (upper panel) and i-K colors (bottom panel), respectively,
for a set of   solar-metallicity PISA isochrones  with ages from  0.5\,Myr  to 15\,Gyr. Different 
colors indicate different effective temperature values of M-type stars from 3000\,K (violet lines)
up to 4000\,K (red lines) with  a step of 50 k.}
\label{loggcolpisa}
 \end{figure}
     \begin{figure}
 \centering
 \includegraphics[width=9cm]{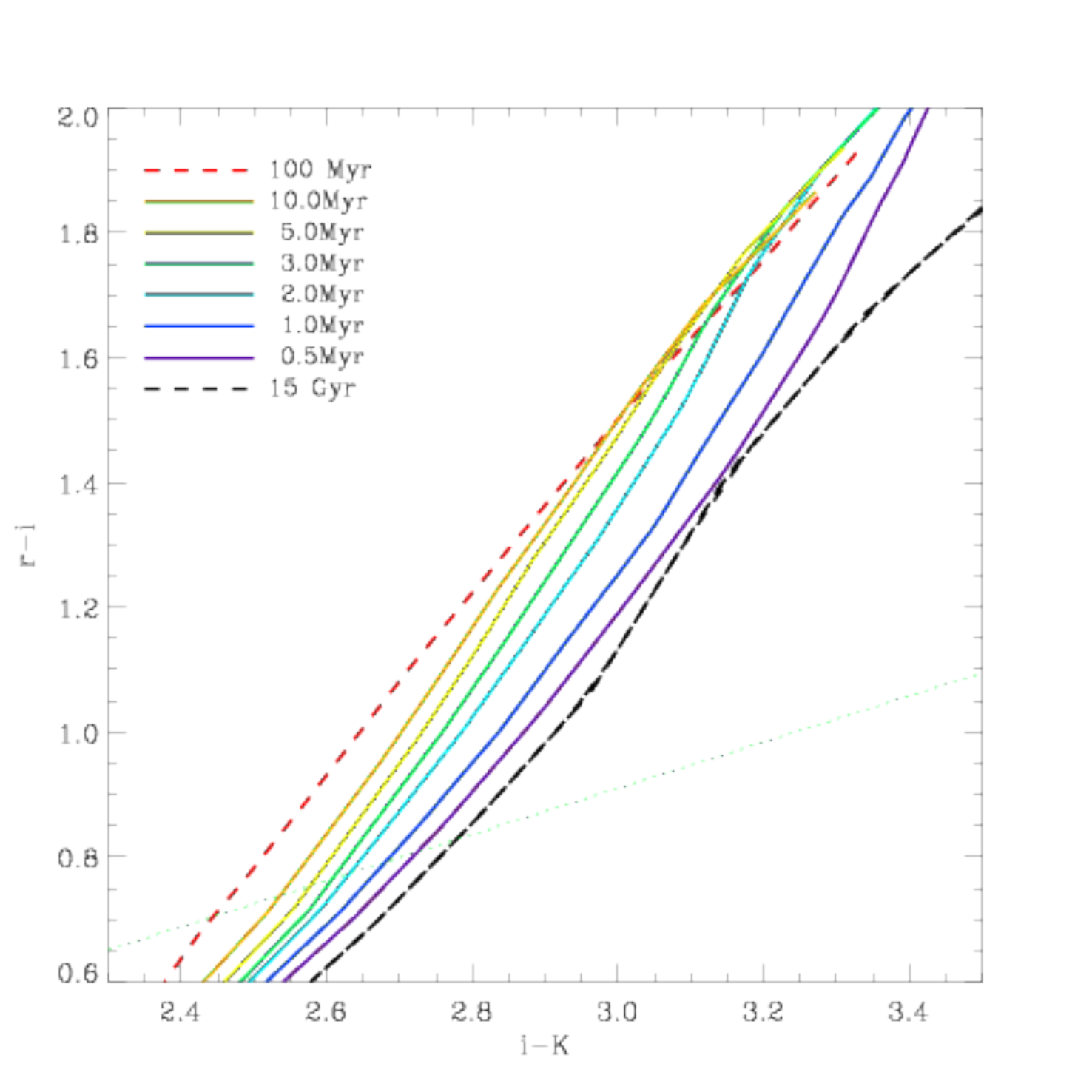}
\caption{Zoom of the r-i vs. i-K diagram showing isochrones of different ages. The dotted green line is the limit
used to select M-type stars.}
\label{riikmselzoom}
 \end{figure}
     \begin{figure*}[!h]
 \centering
 \includegraphics[width=\textwidth]{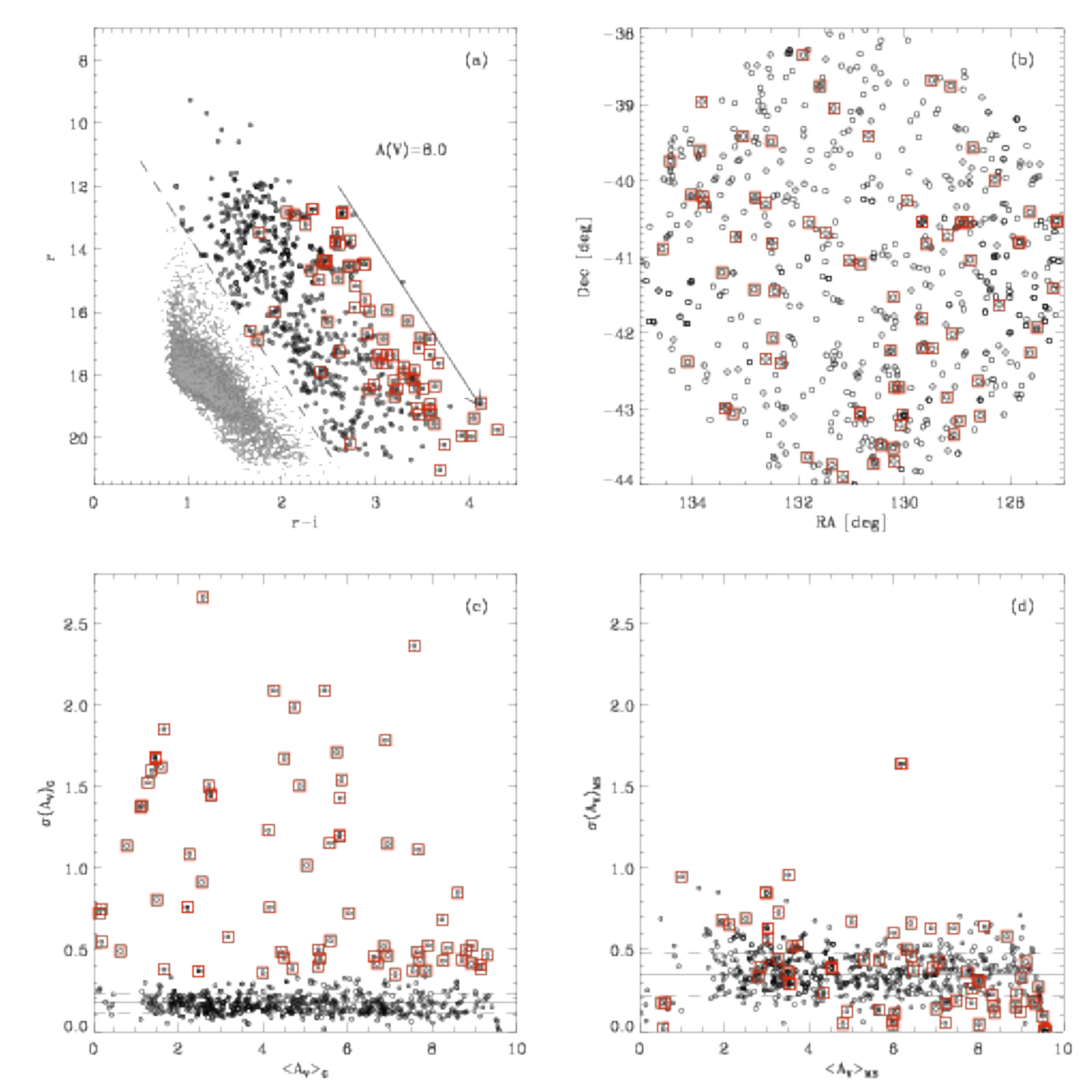}
\caption{(a) r vs. r-i diagram of all the objects selected as M-type stars (grey symbols).
 Black dots are those with r magnitude brighter than
the limit indicated by the dashed line, while red squares indicate the subsample including 
M-type for which the A$_V$ values are not consistent with solar metallicity giants.
 The reddening vector is also indicated. Spatial distribution (b) and $\sigma(A_V)_l$ 
 vs. the $<$A$_V>_l$ values of bright M-stars, obtained assuming the giant locus (c)
  and the MS locus (d).
 The median and the median $\pm$ the pseudo-sigma  are indicated by the solid and dashed lines, 
 respectively,  in the (c) and  (d) panels.}
\label{mselgig}
 \end{figure*}
 
 A further feature of the  M-type stars is  that in the r-i vs. i-J diagram the theoretical 
loci of PMS, MS and giant stars  almost 
 overlap, while in the  r-i vs. i-H and r-i vs. i-K diagrams they are quite  separated.
 This occurs because both i-H and  i-K colors of giant M stars of a given temperature are 
  redder than those of MS stars
as  is clearly evident  in Fig.\,\ref{loggcolpisa},
where the stellar gravities, for a set of  
  theoretical PISA isochrones at solar metallicity,
   are shown as a function of the i-J and i-K colors, respectively.
This feature implies that the  i-H and i-K colors of
 the PMS isochrones of  ages $\gtrsim$10\,Myr are  similar to MS star colors,
 while for the very young PMS stars (0.5\,Myr in the case of our models)
 the i-H and i-K colors are more similar to those of the giant locus,
 as shown in Fig.\,\ref{riikmselzoom}.

  In the r-i vs. i-H and i-K  diagrams, PMS isochrones with ages from 0.5 to 5\,Myr 
  are located between the giant and the MS loci. 
 Therefore the PISA  models predict that  the i-H and i-K colors are sensitive
 to the stellar ages and gravities, at least for Log\,g$>$3.0.
 M-type giants, being very low gravity objects (Log\,g$<$3.0),
  have i-H and i-K colors more similar to YSOs  ($<$1\,Myr) rather than to objects with 
 ages $>$5-10\,Myr.   Therefore, within the limits of the photometric errors,
 at a given temperature (or analogously at a given r-i color), the i-H and i-K colors
   can  be exploited to constrain also 
the luminosity class of M-type stars, as described below.

For each individual M star,
we computed a  reddening value from each of
the three   r-i vs. i-J, i-H and i-K diagrams 
by projecting back the observed colors along the 
corresponding reddening vector onto one of the possible unreddened theoretical 
loci\footnote{The reddening value is converted in absorption by 
following the relations given in Table\,\ref{extinctiontab}}.
 For the objects located bluewards of the related model isochrone,
we computed the A$_V$ only if the color is bluer than
	   the related model isochrone by less than the maximum photometric error, e.i. 0.1414 mag.
	   In these cases the A$_V$ is negative but compatible with A$_V$=0.

 As discussed before, 
 in the r-i vs. i-J diagram, the theoretical loci of  giants, MS and PMS stars  almost overlap, and therefore
the derived absorption values A$_V$   will be almost independent of the gravity.
 On the contrary, in the  r-i vs. i-H or i-K diagrams,  the
 derived A$_V$ will  depend on the gravity,
 since in these diagrams the loci are quite separated.
Therefore, for each star we derive the  A$_V$ sets (A$_V^J$, A$_V^H$, A$_V^K$)$_l$
 from the three diagrams and each adopted evolutionary status l (with l=MS or G, indicating the
 loci of MS stars and giants, respectively, or PMS indicating any isochrone with age $\lesssim$10\,Myr).
%
For each given evolutionary status, we derived the means $<$A$_V>_l$, i.e. 
$<$A$_V>_{MS}$,
$<$A$_V>_{PMS}$,
$<$A$_V>_{G}$ and the standard deviations  $\sigma (A_V)_l$, i.e. 
$\sigma (A_V)_{MS}$,
$\sigma (A_V)_{PMS}$,
$\sigma (A_V)_{G}$ from the three diagrams. 
 For the case of the PMS state, the A$_V$ calculation includes the multiple Age-Dependent sets 
of (A$_V^J$, A$_V^H$, A$_V^K$) corresponding to the isochrones of 0.5, 1.0, 2.0, 3.0, 5.0 and 10.0\,Myr ages. 

Clearly, the $\sigma$ will be large if the adopted evolutionary stage is not the correct one.
Thus,
we define the best A$_V$ estimate of each star as the mean 
 $<$A$_V>_l$ corresponding to the minimum $\sigma (A_V)_l$.

\subsection{Bright M-type population}

Figure\,\ref{mselgig} (a) shows the r vs. r-i diagram of the  M-type stars.
We note that they appear to be grouped in two distinct populations, separated by the dashed line shown in the diagram.
We first considered the brightest group to which we will refer to as bright M-type  (BM) 
population and, in particular, we looked at the  spatial distribution of this sample,
that, as shown in the figure (panel b), does not show any significant clustering.
 From this distribution, we infer that these objects 
are not correlated with the young population studied above, but, on the contrary, they are consistent with field
 M-type MS  or giant stars, homogeneously  distributed over the region.
In order to constrain the luminosity class of these objects,
we consider the values of $\sigma(A_V)_{MS}$ and 
$\sigma (A_V)_{G}$ defined in the previous section. 
 The results are shown in Figure\,\ref{mselgig}  where we show 
the scatter plots of dispersion vs. average value for 
 $<A_V>_G$ and  $<A_V>_{MS}$, in panels c and d, respectively.

The $\sigma(A_V)_G$ data show
 a tail of outliers  at high values (red squares). 
   We estimated 
the statistical parameters of the $\sigma(A_V)_G$ distribution
while minimizing the effect of outliers,
 by computing the median and the pseudo-sigma as the difference between median and the 16$^{th}$ percentile,
 corresponding to 1\,$\sigma$ of a Gaussian distribution
(0.50-(0.68/2.0)=0.16). 
The median of the distribution is 0.17 mag   
while the pseudo-sigma is 0.06 mag.  
On the contrary, the median of the $\sigma(A_V)_{MS}$ is  0.35  
and the pseudo-sigma of 0.13.
The first conclusion is that the values obtained assuming the two models show  very different shapes.
In fact,
the $\sigma(A_V)_G$ distribution  of 
most of the objects (black points) shows  a very narrow  peak,
while  the  $\sigma(A_V)_{MS}$ distribution is much  larger. 
For most of the objects,  $\sigma(A_V)_G<<\sigma(A_V)_{MS}$.
This allows us to deduce that for this sample i) the A$_V$ values obtained in each of the CCD by using 
the giant model are more accurate than  those obtained by using the model of MS stars; ii) 
by computing the mean of the three values associated with the giant model, we obtain not only  more accurate but also more precise
photometric A$_V$ values; iii)  the very narrow peak of the $\sigma(A_V)_G$ distribution of these objects 
suggests that all of them are consistent with a population of giants rather than MS stars.  
This latter result is consistent with their position in the r vs. r-i diagram, since it is more likely that very 
reddened and relatively  bright   objects like these are M-type giants rather than M-type MS stars.

Finally,  we selected 76  objects 
 with  $\sigma(A_V)_G>0.35$  (i.e. larger than 3\,pseudo-$\sigma$ from
the median of the distribution) and for which $\sigma(A_V)_{MS}<\sigma(A_V)_G$. 
For these objects, 
 the quite large spread of the $\sigma(A_V)_{MS}$ values does  definitely suggest
 they are not MS stars.
 We suspect that the $A_V$ estimates obtained for these objects might not be
reliable  because we used only solar metallicity models, that are not appropriate if they are 
  M-type stars with non-solar metallicities. 
  As a consequence, with the models adopted in this analysis, we are not able 
 to classify these 76 objects.

In conclusion, from the statistical values, indicated in Figure\,\ref{mselgig}, 
 we deduce that most of  the bright  M-type stars (551/627)  are giants, 
 affected by absorption A$_V$ ranging from $\simeq 1.5$ to 10,
with an error on  A$_V$  of  about 0.17 
(corresponding to the 
median of the    $\sigma(A_V)$ values).

     \begin{figure}
 \centering
  \includegraphics[width=9.0cm]{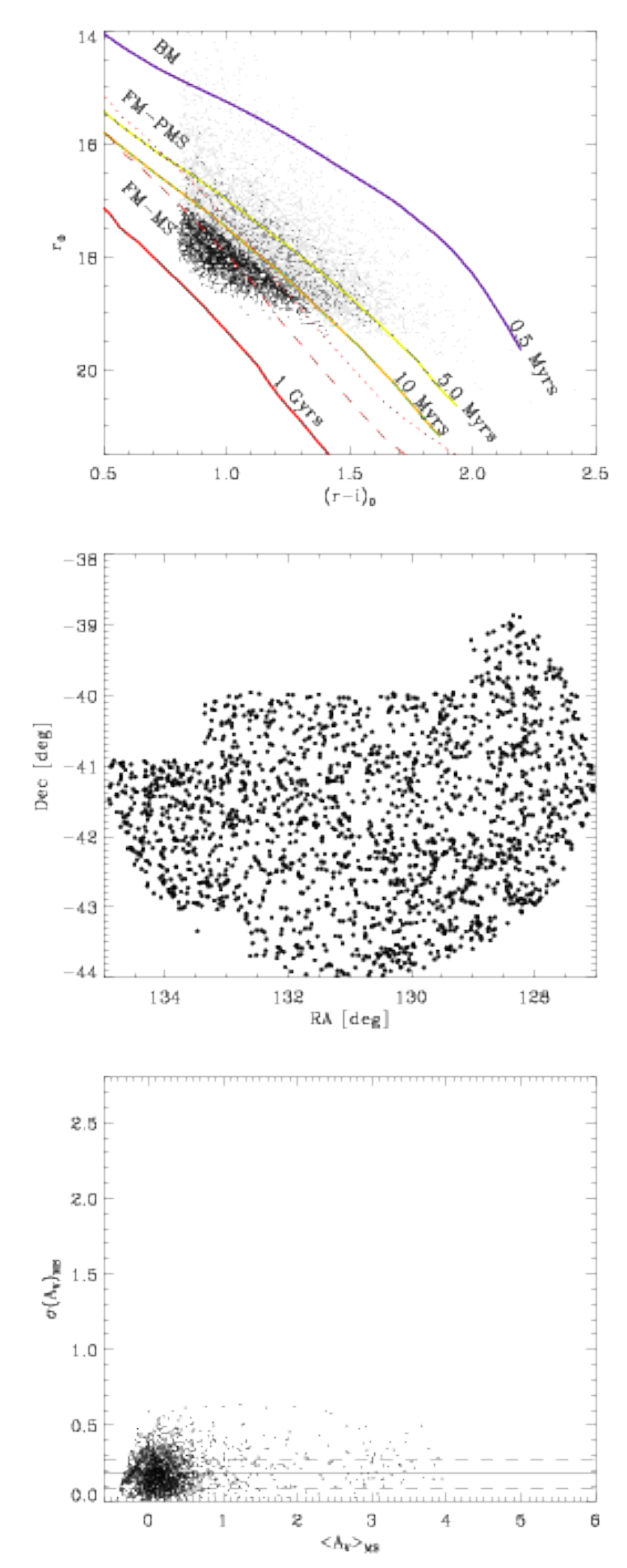}
\caption{r$_0$ vs. (r-i)$_0$ diagram of the faint M-type population (upper  panel). Black dots are those 
selected as MS foreground field stars (see text).
Solid lines are the  0.5, 5.0, 10.0\,Myr and the 1\,Gyr isochrones at the cluster distance.
Dotted and dashed lines represent the 1\,Gyr isocrone at 300 and 400\,pc, respectively.
 Spatial distribution (middle panel) and $\sigma(A_V)_{MS}$  (bottom panel) of the selected MS foreground field stars.
  The median and the median $\pm$ pseudo-$\sigma$ 
of this distribution
 are indicated as the solid and dashed lines.}
\label{mselmsr0ri0}
 \end{figure}
     \begin{figure}
 \centering
 \includegraphics[width=9.0cm]{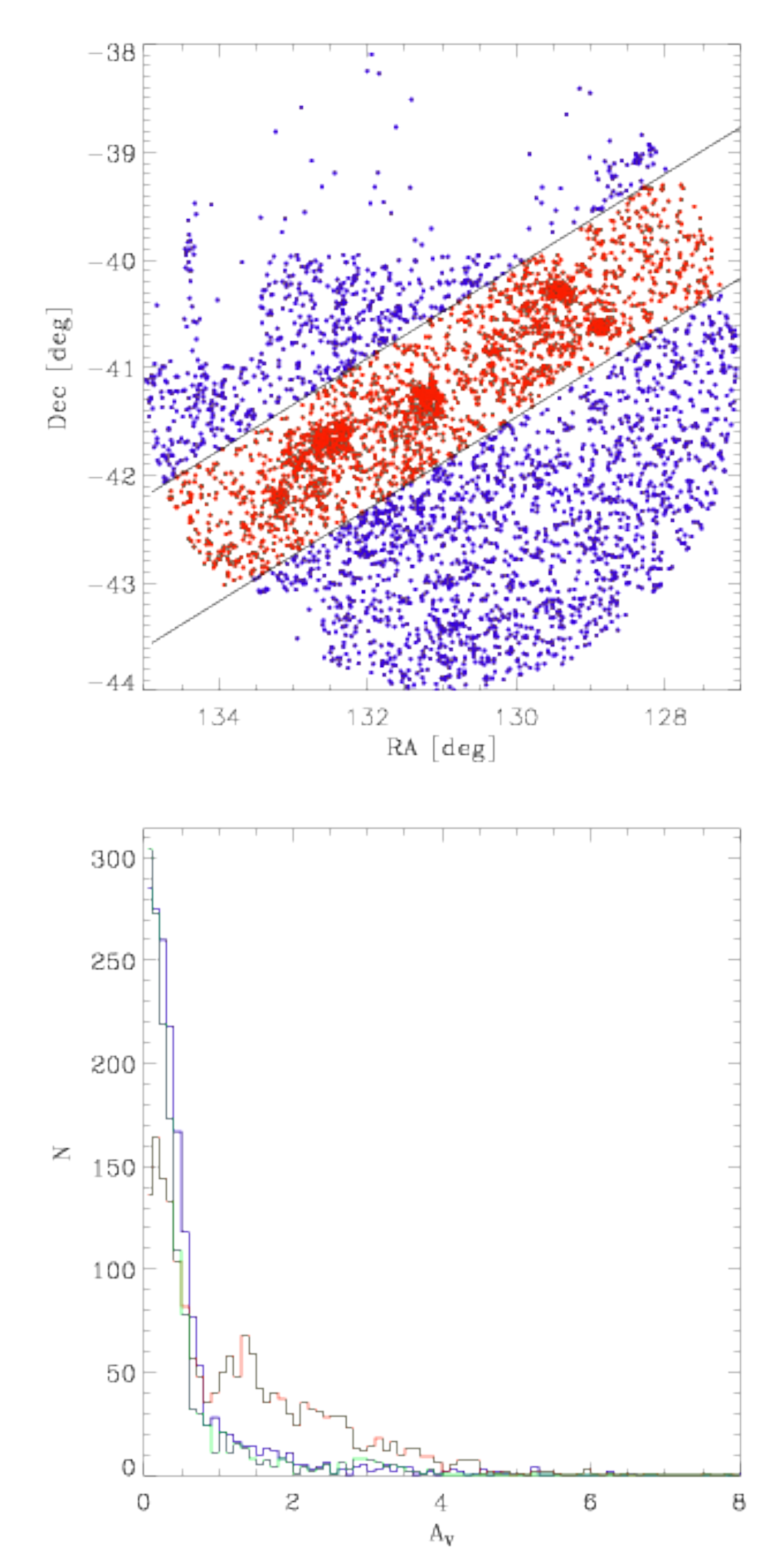}
\caption{Upper panel: spatial distribution of the FM-PMS population, divided in two samples, internal (red points) and external
(blue point) to the strip arbitrarily defined by the two solid lines. Bottom panel: histograms of the A$_V$ distribution
obtained by using
the FM-PMS population within the strip (red histogram), FM-PMS population outside the strip (blue histogram) and 
the MS foreground field stars (green histogram).}
\label{avbreak}
 \end{figure}

    \begin{figure*}
 \centering
 \includegraphics[width=\textwidth]{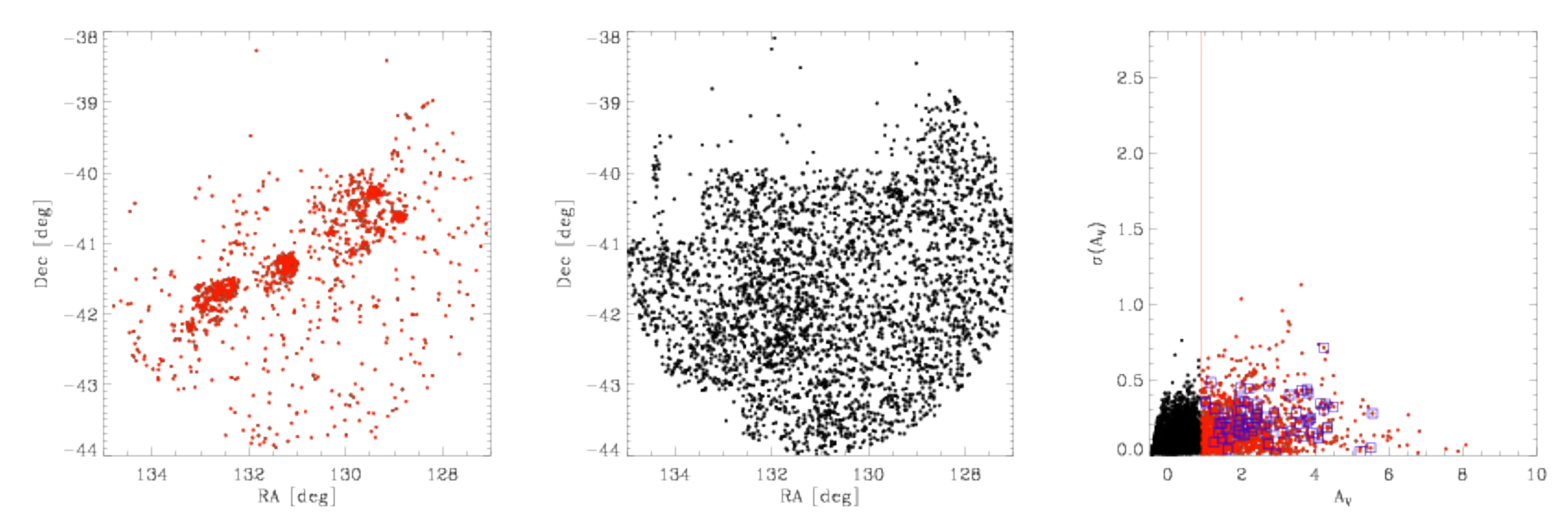}
\caption{Spatial distributions (left and middle panels) and  $\sigma(A_V)$ vs. $A_V$ (right panel)
associated with each star
belonging to the red-FM-PMS (red dots) and near-FM-MS (black dots) populations. Blue squares are the
objects with disk selected from IR and/or H$\alpha$ excesses.}
\label{mselfmpms}
 \end{figure*}
  \begin{figure}
 \centering
 \subfloat{\includegraphics[width=7.5cm]{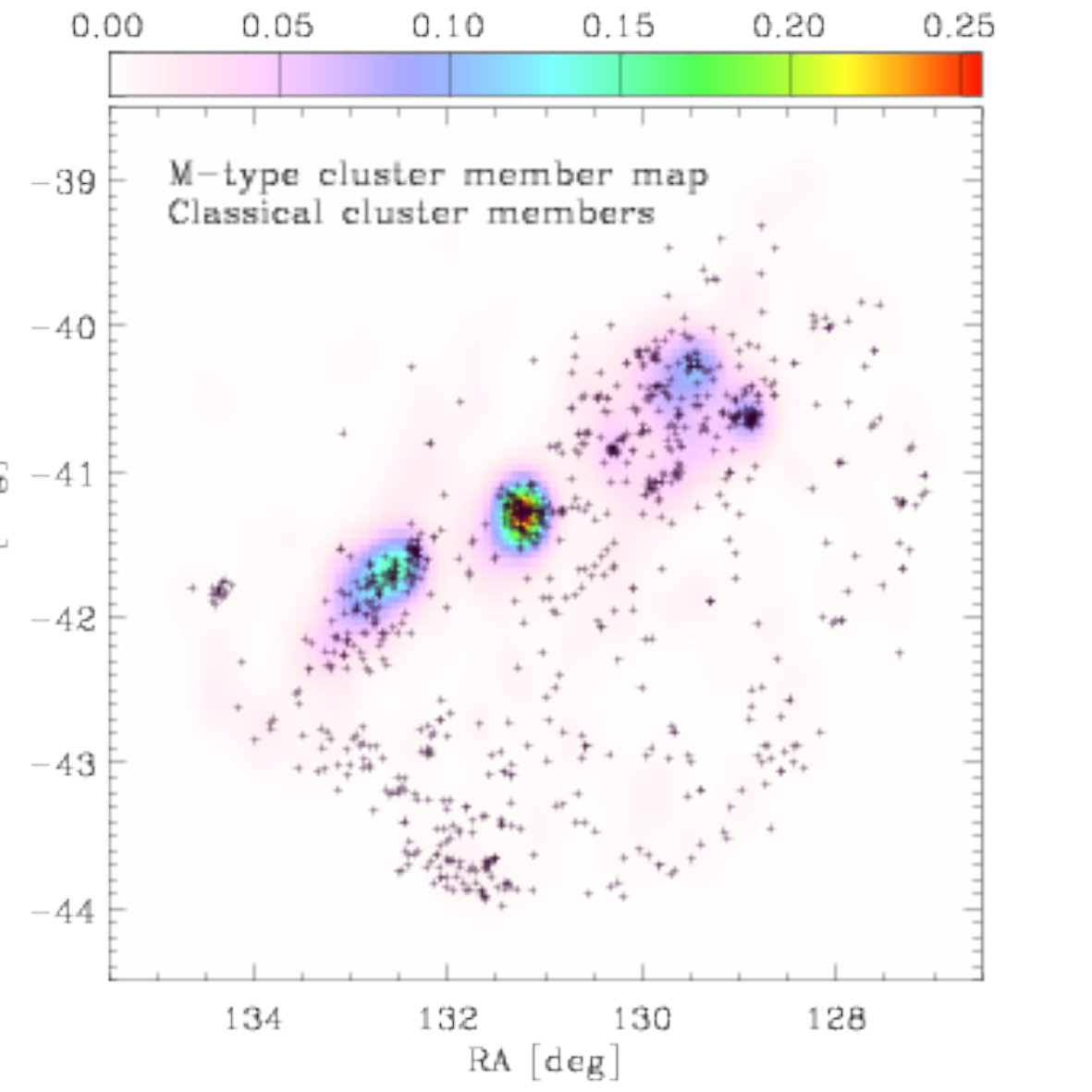}}\\
 \subfloat{\includegraphics[width=7.5cm]{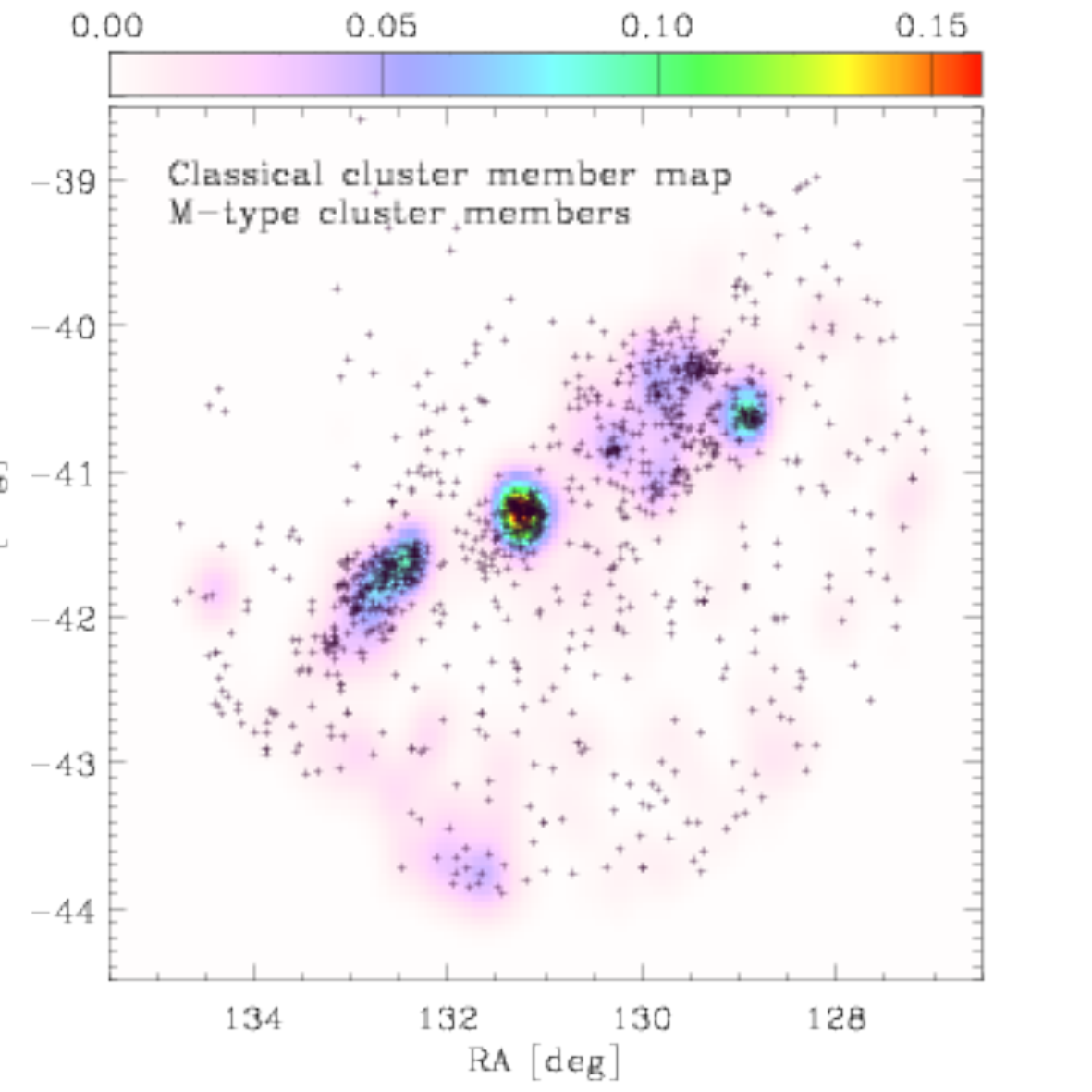}}\\
\caption{Upper panel: spatial distribution 
of  cluster members selected as M-type 
PMS stars in Sect.\,\ref{fmpms}, plotted as a  two-dimensional histogram
smoothed with a Gaussian Kernel. 
The density scale is the number of stars per bin and per arcmin$^2$.
Overplotted black crosses are candidate members selected with 
classical methods. 
Lower panel: spatial distribution 
of  cluster members selected with classical methods, plotted as a  smoothed two-dimensional histogram.
 Overplotted black crosses are  cluster members selected as M-type 
PMS stars.}
\label{mselmempmsmap}
 \end{figure}
 %

\subsection{Faint M-type population}
The faint M-type (FM) population includes all  objects
selected as M-type stars, with r magnitudes fainter than the limit indicated by the dashed line in 
 panel (a) of Fig.\,\ref{mselgig}, shown in the figure with grey symbols.

We expect this population to include  field stars as well as  YSOs belonging to the clusters
we are studying here. Therefore, for each star we computed the 
triplets (A$_V^J$, A$_V^H$, A$_V^K$)$_{MS}$,
(A$_V^J$, A$_V^H$, A$_V^K$)$_{G}$ and
(A$_V^J$, A$_V^H$, A$_V^K$)$_{PMS(t)}$, 
with t indicating the isochrone ages 0.5, 1.0, 2.0, 3.0, 5.0 and 10.0 \,Myr.

 As done for the BM population, 
 for each triplet, we computed the mean  $<$A$_V>_l$ and $\sigma (A_V)_l$ 
and we considered the best A$_V$ estimate for a given star
 the $<$A$_V>_l$ for which the $\sigma (A_V)_l$  is  smallest.

This condition allows us to derive an  A$_V$  more accurate 
than the one  obtained by using only one diagram that could be affected
by the possible presence   of circumstellar disk, with excesses in one or more of the
three  J, H or K bands \citep{ciez05}.
We note that if the M-type stars are affected by strong IR excesses, their colors
are redder than the limit adopted to select M-type stars and, therefore, they are not included
in the sample of M-type stars. 
Nevertheless, young M-type stars with small IR  excesses
can fall in the region  of M-type stars. In these cases,
 the A$_V$  can be affected  by the IR excesses and 
the $\sigma (A_V)$ could be larger than the one derived  for disk-less M-type stars.
 These cases can be considered as candidates M-type young 
 stars with circumstellar disk.



    
\subsubsection{Faint M-type MS stars}
As already mentioned, due to the partial overlap of the oldest PMS isochrones  
with the MS locus and of the youngest 
 ones with the giant locus,  PMS stars  can be misclassified as MS stars or giants and vice-versa.
Therefore, in case of star forming regions, 
it is not possible to definitively assign the age to the objects.
Nevertheless, 
the best A$_V$ values obtained with this method, within the uncertainties due to the photometric errors and the model
accuracy, can be considered reliable to obtain intrinsic colors and magnitudes, since they do
 not depend on the knowledge of the
stellar evolutionary stage but only on the photometric indices.

Fig.\,\ref{mselmsr0ri0} shows the unreddened r$_0$ vs. (r-i)$_0$ diagram, 
derived from the best A$_V$ values, for the faint M-type population, compared with
 the  0.5, 5.0, 10.0 Myr and the 1\,Gyr  isochrones at the cluster distance.
 In this diagram,  the only unknown parameter is the distance but, if we assume that the young population
 can be found only between the 0.5 and 10\,Myr isochrone at the distance of 750\,pc 
 (see Sect.\,\ref{kinematic}),
 we can conclude that all the objects with  r$_0$ larger than the 10.0\,Myr isochrone
 at the cluster distance (to which we added a delta r equal to 0.2 to take into account the errors) 
 are not compatible with YSOs belonging to our clusters and therefore
 are expected to be field stars. 
 Based on the position of the 1\,Gyr isochrone,  at the cluster distance, we deduce they cannot be 
 background M-type MS stars but, instead, they are compatible with foreground  M-type MS stars with distance smaller
  than the cluster distance
 and approximately larger than about 300-400 \,pc. In fact, the 1\,Gyr MS isochrone at 300\,pc (400\,pc) approximately corresponds
 to the M-late  (M-early) lower limit in r$_0$ that  we adopted to delimit this sample of field stars. 
 Our conclusion is confirmed by 
 their spatial distribution, that is quite uniform, as expected for field foreground stars.\footnote{M-type
  MS stars at distance smaller than 300-400\,pc are, on the contrary, expected to be brighter and 
 therefore  overlapping with the PMS region at the cluster distance.} We will refer to this population as the faint M-type MS (FM-MS)
 stars.
 
Fig.\,\ref{mselmsr0ri0} shows also the $\sigma (A_V)_{MS}$ values as a function of the $<$A$_V>_{MS}$.
As done for the giants of the bright M-type population, we computed the median and pseudo-$\sigma$ of
the  $\sigma (A_V)_{MS}$ values
that are found to be, respectively, 0.18, equal to that of bright giants,  and 
 0.1, larger than the analogous value found for giants (0.06). The larger dispersion derived for MS stars may
 be  due to the very small values 
of A$_V$  ($\lesssim$0.8) found for most of these stars.
In conclusion, we selected  1869 faint M-type objects that we classify as MS foreground field  stars
with distances between 300-400 and $\sim$750\,pc.  

\subsubsection{Faint M-type PMS stars\label{fmpms}}
In the following, we will analyse  the 
 faint M-type population in the PMS photometric region (FM-PMS), 
 i.e. that included between the 0.5 and 10\,Myr isochrones at the cluster distance.
 As already mentioned,  this  FM-PMS population is expected 
 to include the M-type members of the clusters but also a fraction of 
contaminating field stars at distance smaller than about  300-400\,pc.  The spatial
distribution of this sample, shown in Fig.\,\ref{avbreak}, 
shows several clumps of stars, with a pattern  
 very similar to the one found for the YSOs selected with {\it classical} methods
  (see Fig.\,\ref{radecirex}),
  and this suggests 
 that the genuine M-type YSOs are mainly distributed within the spatial strip
drawn in the upper panel of Fig.\,\ref{avbreak} with red symbols. On the contrary, the spatial distribution of the 
objects outside the strip is quite homogeneous. Therefore, as first approximation,
  we assume that the FM-PMS within the 
strip is dominated by the young cluster members, while the FM-PMS  outside the strip
is dominated by  the field star population.
Figure\,\ref{avbreak} shows also the histograms of the A$_V$ values for these two populations 
and for  the MS foreground field stars. 
It is evident that
the population dominated by the young cluster members shows two peaks, centered at about A$_V\sim$0.3 and A$_V\sim$1.5
while both  the other two populations, dominated by the field stars,  are very similar and 
shows a single peak at about A$_V$=0.3, with a rapid decrease for 
A$_V\gtrsim1$. The A$_V$ distribution of the young cluster members
shows  a minimum at A$_V=$0.9, corresponding to a sharp  drop in the two field distributions.
This is the A$_V$ limit that clearly separates the young population from the field population.
We cannot exclude that
a small fraction of YSOs might be affected by reddening smaller than A$_V=$0.9
and, vice-versa, a small fraction of MS field stars might be affected by A$_V>$0.9, like the cluster
stars. Nevertheless,  the inclusion of this very small fraction of contaminants and the loss
of potential few reddened YSOs do not affect our results.   

Based on this finding, we split the FM-PMS population  in two subsamples,
one including objects with A$_V>$0.9 and the other including the objects with A$_V\le $0.9, respectively,
 to which we will refer as red-FM-PMS
and  near-FM-MS populations. 

Figure \,\ref{mselfmpms} shows the spatial distributions of the red-FM-PMS and the  near-FM-MS populations.
 The spatial pattern of the red-FM-PMS population
is very similar to that of members selected  with other membership criteria
(see Fig.\,\ref{radecmem}) and therefore we deduce that the large majority of 
these objects are genuine members of the three young clusters
in  RCW\,33, RCW\,32 and RCW\,27. As found  by using other methods, there are still candidate members
sparsely distributed in the regions outside. Even though  they could be less-likely members, we do not attempt to discard them 
since the shape of the clumps with high density of cluster members is very irregular and we cannot discard 
the hypothesis that many of them are true YSOs. Therefore, we keep all objects of the red-FM-PMS sample 
as M-type candidate cluster members.
We note that the sample of 907  cluster members, previously selected with classical methods,
 includes  only 84 M-type stars.
Among these, 72 (8) belong to the  red-FM-PMS (near-FM-MS) populations and are therefore confirmed cluster members.
In fact, based on the A$_V$ distribution, we know that a small fraction of cluster members is expected to fall 
also in
the near-FM-MS population, mainly populated by non members.
 Only 3 objects belong to the population  of foreground stars and one to the BM population.
 
A very different spatial distribution is, instead, found for the near-FM-MS population. It is quite uniform, even though 
there is a slightly higher stellar density around 
 RA=131\fdg8 and Dec=-42\fdg4,
that corresponds to the nominal centre of the Trumpler\,10 association \citep{khar13}.
The result is better visible 
if we restrict the analysis to objects within a 0.5\,mag
strip around a 20\,Myr isochrone at the distance of Trumpler\,10 (422\,pc) found with the PM analysis. 
However, since the spatial distribution of Trumpler\,10 is very sparse, 
in this work we do not attempt to select its members. Therefore,
we classify all the objects belonging to the near-FM-MS population as objects not associated with the three young clusters.

In summary, we have selected 
627 objects 
belonging to the M-type bright population and 6\,462
 objects of the M-type faint population,
for which we derived the individual interstellar reddening.
Among the objects in the faint population, there are 
1\,869 foreground field MS  stars with distance larger than 300-400\,pc, 
3\,393 near-FM-MS stars, within the PMS photometric region, affected by small reddening that are not correlated to  the young clusters and finally
1\,200 FM-PMS stars, that we consider as M-type candidate members of the young clusters. 

Figure \,\ref{mselfmpms} shows  also the $\sigma(A_V)$ as a function of A$_V$ for 
the red-FM-PMS and the near-FM-MS samples.
In this figure, we highlighted the  YSOs 
with IR  and/or H$\alpha$ excesses, selected also as M-type stars.
We find that very few of them show a $\sigma(A_V)$ significantly larger than the other objects, 
while the remaining
show a $\sigma(A_V)$ distribution very similar to that of the other members. This result suggests that, 
for the  M-type members,
the effects due to the presence of circumstellar disk on the   A$_V$ values derived with this method 
can be considered negligible. 
\section{Discussion}
\subsection{Spatial distribution analysis}
The large number of likely M-type PMS stars selected in this work 
are consistent with results 
 recently discussed in \citet{dami18} and are related to the higher intrinsic brightness 
 of PMS M-type stars, with respect to MS M-type  stars. 
This implies that at a given mass, a 1\,Myr old M star can be detected at a distance
much larger than a MS M star \citep{dami18}. 
This is a very important property of M-type stars,
that has been little exploited up to now in the literature and that, instead, can be used as  a very efficient 
tool to detect the low mass  population of YSOs 
in  star forming regions, dominated by M-type stars. In addition,
 by deriving  the individual reddening values   
using different optical-IR color combinations and the model predictions  for different ages,
we are able to separate  MS and  giant M-type stars, mainly  expected to be found 
respectively, at small and large distances from us, and to recognize M-type YSOs.

This result is confirmed from the spatial distributions shown in Fig.\,\ref{mselmempmsmap},  
 for the M-type young cluster members,
and for the members selected with classical methods. The coloured maps are two-dimensional histograms 
to which we applied a smoothing  with a 
Gaussian kernel. 

The comparison of the two spatial distributions  shows
a clear spatial correlation between the YSOs
selected with classical methods and the M-type candidate members selected in Sect.\,\ref{fmpms},
confirming the peculiar patterns of the three regions,
with a more concentrated distribution of  YSOs in Cr\,197 (RCW\,32). 
The spatial distribution of  objects in the cluster Vela T2 (RCW\,33) is slightly more sparse with an ellipsoidal shape,
while that in Vela T1 (RCW\,27) is very sparse with evidence of several clumps.   
 
 We verified that
the spatial distribution of the faint FM-MS population  and that of the  objects selected as
 near-FM-MS  and BM (background) giants  are  not 
 indicative of any significant stellar density clumps, but rather are consistent with 
  a uniform and homogeneous distribution.

\subsection{SF history of M-type stars}
    \begin{figure*}
 \centering
 \includegraphics[width=\textwidth]{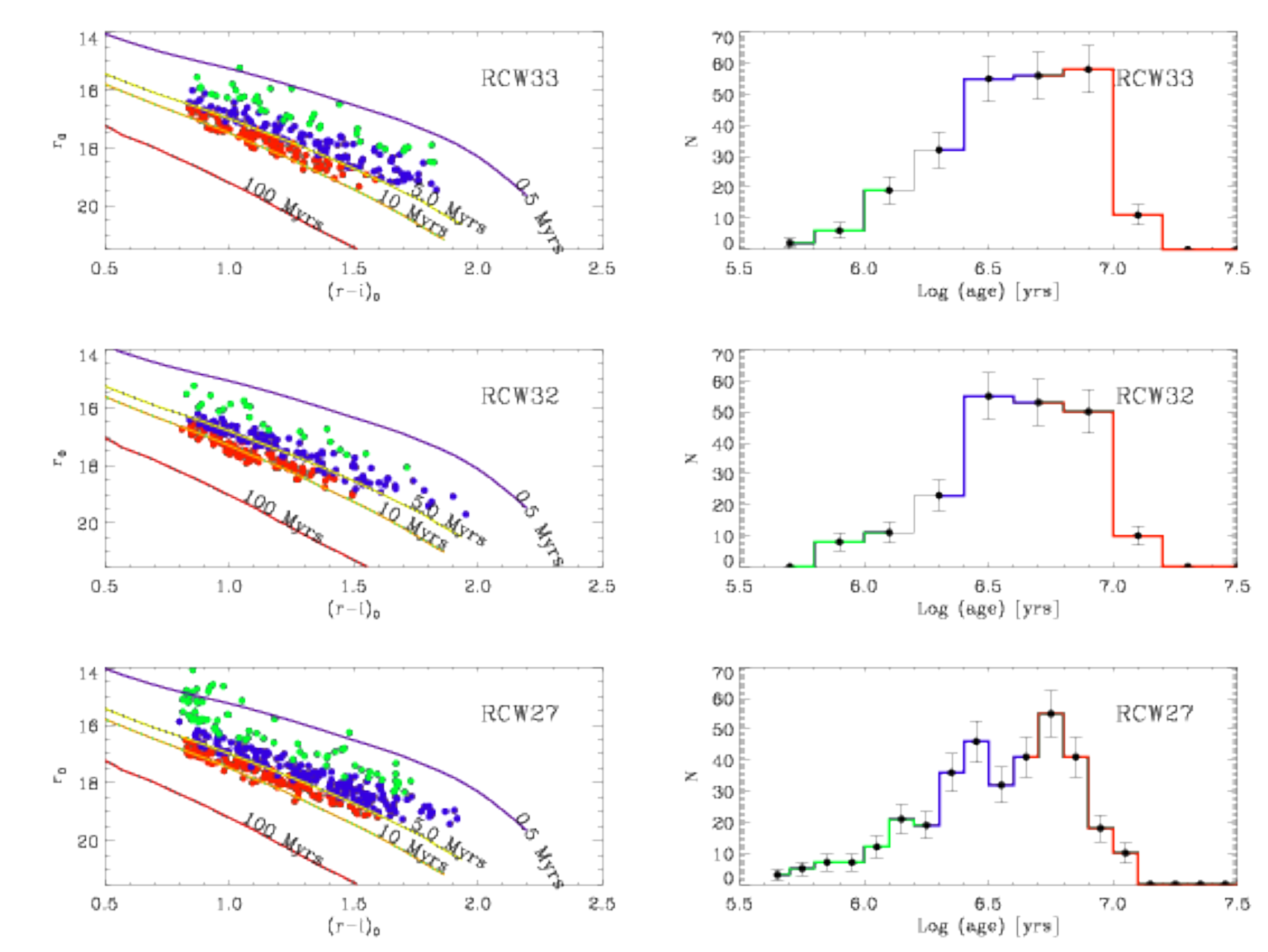}
\caption{Intrinsic CMDs  and age distributions of the M-type candidate members  falling in the three
regions defined in Table\,\ref{galboxes} and grouped in three age ranges, i.e. t$<$2 My, 2$<$t/[Myr]$<$5 and 
t$>$5 Myr, indicated with green, blue and red symbols, respectively. Several representative isochrones are 
overplotted in the CMD.}
\label{r0ri03reg}
 \end{figure*}

  \begin{figure*}
 \centering
 \includegraphics[width=\textwidth]{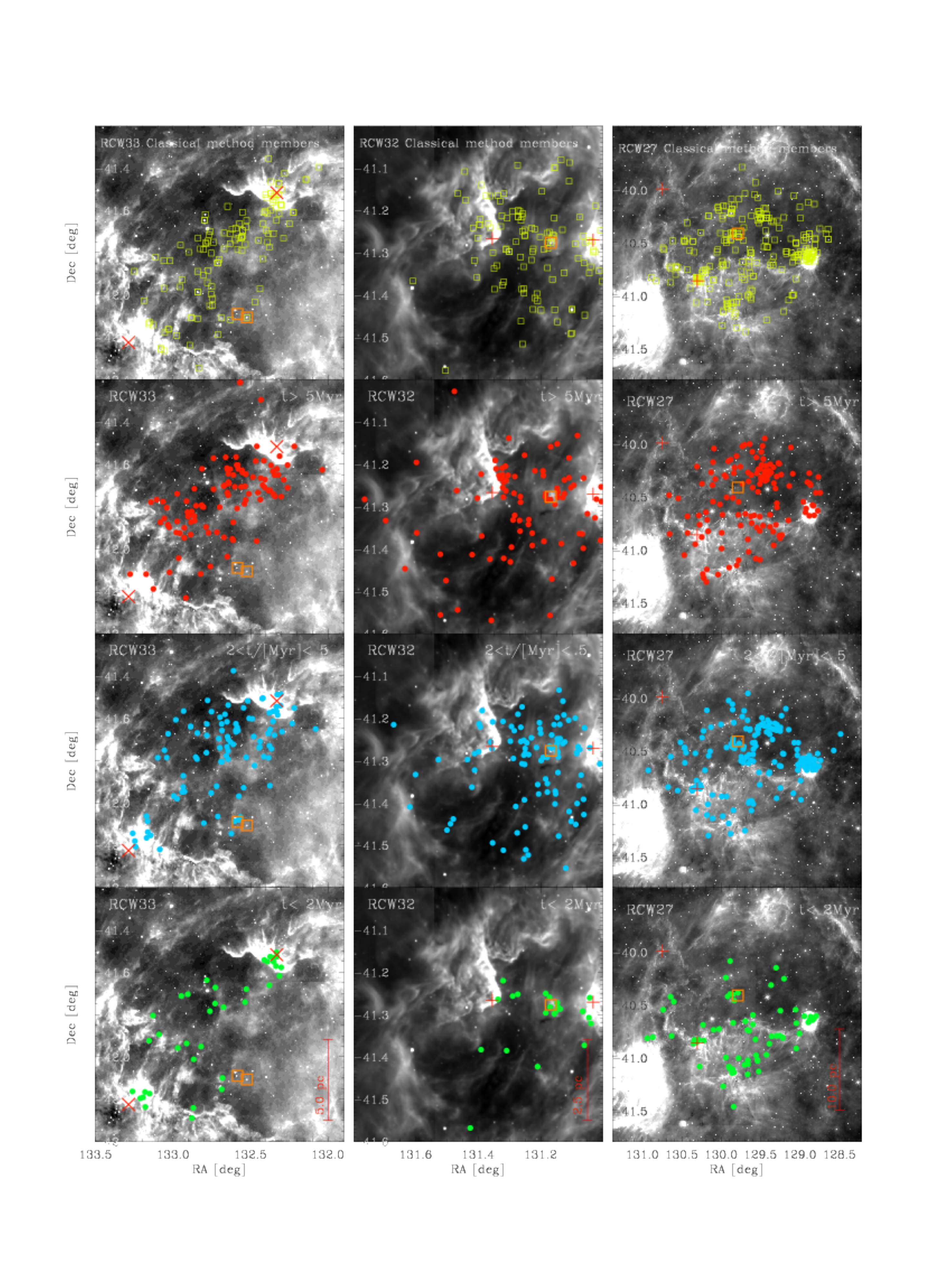}
\caption{Different scale WISE images at 12\,$\mu$m in RCW\,33, 32 and 27 in left, middle and right panels, 
respectively. YSOs selected with classical methods (upper panels) and M-type YSOs in different age ranges
(lower panels) overplotted with yellow squares and red, blue and green points, respectively.
Orange empty squares indicate the position of the ionizing stars, while red crosses and plus symbols 
correspond to C$^13$O clouds 14 and 15 and to SFO sources, respectively.}
\label{magebinimg}
 \end{figure*}
For the M-type YSOs, we computed stellar ages  
assuming the distance d=750\,pc 
and using the individual reddening values derived in Sect.\,\ref{mtypestars}. 
A minimum and a maximum value to stellar ages have been estimated 
by considering the photometric errors on r and i and the error on A$_V$.
The A$_V$ and stellar ages of M-type members together with their coordinates and optical/NIR photometry 
are given in Table\,\ref{emmetab}.
\begin{table*}
\caption{Optical/NIR photometry, A$_V$ and ages of M-type members.
 Full table available in electronic format only.
 \label{emmetab}}
\centering
\begin{tabular} {c c c c c c c c c c c c }  
\hline\hline
CNAME & RA  &   Dec          & r & i &  J & H & K & A$_V$ & Log(t) & Log(t)$_{\rm min}$  & Log(t)$_{\rm max}$ \\
      & (J2000) & (J2000)    &   &   &    &   &   &       &   (yrs)     &       [yrs]              &     [yrs] \\
\hline
08475118-3928303 &    131.96325 &   -39.47508 &   14.90 &   13.87 &   11.73 &   10.99 &   10.72 & 1.13$\pm$ 0.08 & 5.00 & 5.00 & 5.04\\
08572585-4026000 &    134.35771 &   -40.43334 &   15.85 &   14.77 &   12.59 &   11.78 &   11.50 & 1.28$\pm$ 0.04 & 5.62 & 5.52 & 5.67\\
08575042-4032562 &    134.46010 &   -40.54893 &   15.06 &   14.00 &   11.86 &   11.06 &   10.81 & 1.16$\pm$ 0.02 & 5.00 & 5.00 & 5.00\\
08472297-3816088 &    131.84571 &   -38.26911 &   15.27 &   14.07 &   11.71 &   10.79 &   10.49 & 1.62$\pm$ 0.05 & 5.00 & 5.00 & 5.00\\
08363711-3824341 &    129.15461 &   -38.40948 &   15.58 &   14.51 &   12.41 &   11.58 &   11.36 & 0.91$\pm$ 0.01 & 5.53 & 5.50 & 5.56\\
08372719-4153363 &    129.36330 &   -41.89341 &   14.55 &   13.41 &   11.04 &   10.27 &    9.91 & 1.69$\pm$ 0.12 & 5.02 & 5.00 & 5.15\\
08372719-4153363 &    129.36330 &   -41.89341 &   14.55 &   13.41 &   11.04 &   10.27 &    9.91 & 1.69$\pm$ 0.12 & 5.02 & 5.00 & 5.15\\
08372719-4153363 &    129.36330 &   -41.89341 &   14.55 &   13.41 &   11.04 &   10.27 &    9.91 & 1.69$\pm$ 0.12 & 5.02 & 5.00 & 5.15\\
08360090-3941072 &    129.00376 &   -39.68533 &   15.40 &   14.05 &   11.28 &   10.35 &    9.95 & 2.74$\pm$ 0.06 & 5.04 & 5.00 & 5.10\\
08384800-4202113 &    129.69999 &   -42.03646 &   15.69 &   14.56 &   12.20 &   11.59 &   11.32 & 1.35$\pm$ 0.36 & 5.19 & 5.06 & 6.27\\
08310898-3926229 &    127.78741 &   -39.43969 &   16.06 &   14.82 &   12.35 &   11.49 &   11.17 & 1.93$\pm$ 0.02 & 5.03 & 5.00 & 5.11\\
\hline
\end{tabular}
\end{table*}

Figure\,\ref{r0ri03reg} shows the intrinsic r$_0$ vs. (r-i)$_0$ diagrams and the age distributions
 of the members within 
the  three regions 
with the highest density, defined in Table\,\ref{galboxes}. Objects have been distincted in 
three age ranges, i.e. t$<$2 Myr, 2$<$t/Myr$<$5 and  t$>$5 Myr.
\begin{table}
\caption{Galactic coordinates of the boxes around the three \hii regions with the highest
star density  \label{galboxes}}
\centering
\begin{tabular} {c c c c c}  
\hline\hline
\hii & Min. l & Max. l& Min. b & Max. b\\
\hline
RCW33 & 262.0 &263.2 &  1.2 &  1.8\\
RCW32 & 261.2 &261.9 &  0.6 &  1.3\\
RCW27 & 259.5 &261.1 & -0.2 &  1.2\\
\hline
\end{tabular}
\end{table}

The age distributions of the three clusters show  a similar  spread with ages   from 0.5 to about 13\,Myr, but the
population of RCW\,27 seems to contain a double generation  of YSOs, one with median age 
about 2-3\,Myr and another with median age about 5-6\,Myr.
Figure\,\ref{magebinimg} shows the spatial distributions of the 
YSOs 
selected with the classical methods and the
M-type YSOs distincted in the same
three age ranges defined above,
 compared with the dust distribution shown by the IR WISE 
images at 12$\mu$m, in the three regions with the highest density. 
The position of the ionizing   stars and known  clouds  in the region
are given in Table\,\ref{starclouds}.
\begin{table}
\caption{Coordinates of the ionizing stars and of known clouds in the region.
 \label{starclouds}}
\centering
\begin{tabular} {c c c c}  
\hline\hline
\hii & ID & RA [deg] & Dec [deg] \\
\hline
RCW33 & HD75759 &132.58 &-42.09\\
RCW33 & HD75724 &132.52 &-42.11\\
RCW32 & HD74804 &131.17 &-41.28\\
RCW27 & HD73882 &129.79 &-40.42\\
RCW33 & Cloud 14 &132.33 &-41.52\\
RCW33 & Cloud 15 &133.29 &-42.23\\
RCW32 & SFO58 &131.36 &-41.27\\
RCW32 & SFO57 &131.03 &-41.27\\
RCW27 & SFO56 &130.75 &-40.00\\
RCW27 & SFO55 &130.30 &-40.87\\
RCW27 & NGC2626 &128.88 &-40.67\\
\hline
\end{tabular}
\end{table}

 For each of the three clusters, Vela T2, Cr\,197 and Vela T1, and 
following the \citet{fasa87} approach, we applied the
Kolmogorov-Smirnov test on two samples and two dimensions, 
to the spatial distributions of each pair of subgroups of YSOs with different age.
We determined the probability of them being extracted from the same parent distribution,
as reported in Table\,\ref{pvalues}, where the {\it p}-values are given for each region.    
We discuss these probabilities in the following sections.
\begin{table}
\caption{{\it p}-values resulting from applying a two-sample K-S test to the spatial distributions of each pair
of subgroups with different age.
 \label{pvalues}}
\centering
\begin{tabular} {c c c c}  
\hline\hline
\hii & Age range 1 & Age range 2 & {\it p}-values \\
     & [Myr]       & [Myr]       &   \\
\hline
RCW\,33 & [0-2]  & [2-5]  & 0.05 \\
RCW\,33 & [2-5]  & [5-13] & 0.5  \\
RCW\,33 & [0-2]  & [5-13] & 0.04 \\
RCW\,32 & [0-2]  & [2-5]  & 0.2 \\
RCW\,32 & [2-5]  & [5-13] & 0.15  \\
RCW\,32 & [0-2]  & [5-13] & 0.06 \\
RCW\,27 & [0-2]  & [2-5]  & 0.00006\\
RCW\,27 & [2-5]  & [5-13] & 0.04  \\
RCW\,27 & [0-2]  & [5-13] & 0.00006 \\
\hline
\end{tabular}
\end{table}

\subsubsection{Vela T2 and RCW\,33}
The WISE image  shows two bright clouds, corresponding to the  C$^{18}$O clouds 14 and 15
of \citet{yama99a} and several bright rimmed clouds (BRC).
The two ionizing   stars 
 HD75759 and HD75724 lie  in the southern region. The first one is a binary 
with spectral type O9V+B0V, while the second is a  B2III star.
The first generation of stars (t$>$5Myr) are found along the direction defined by 
clouds 14 and  15, with 
a higher concentration  of stars between the two ionizing   stars and  cloud\,14.
The second generation of stars (2$<$t/Myr$<$5)  are mostly found in the same region
 following the same pattern while the youngest stars  (t$<$2 Myr) mostly formed 
two clumps around clouds\,14 and 15, with few  more stars in the middle region, that seems to follow the pattern
defined by the dust density enhancements.  The difference between the spatial distributions of the 
subgroups is not significant or marginally significant with {\it p}-values  comprising between 0.04 and 0.5.
\subsubsection{RCW\,32 and Cr\,197}
 In the region of RCW\,32, cluster members are found
 between the SFO\,58 and SFO\,57 BRCs \citep{sugi94}, 
  associated with the B0\,V-B4\,II ionizing star HD\,74804
 \citep{urqu09}. Both SFO57 and SFO58 are included in the  sample of  candidates for triggered
 SF    \citep{urqu09}, associated to the formation of a photon-dominated region.
Our results show that the first generation of stars formed a clump  around HD\,74804
and another group on the western side, along the direction of SFO\,58 and  attached to it, while the 
second generation formed in the middle region between the two BRCs. The few stars formed in the last
2\,Myr are concentrated around HD\,74804 and in the proximity of the two BRCs. 
 However, the difference between the spatial distributions of the 
subgroups is not significant  with {\it p}-values  comprising between 0.06 and 0.2.
\subsubsection{Vela T1 and RCW\,27}
In the region of RCW\,27, the first generation of stars are found in 
a clump  around  (RA, Dec)=(129.25,-40.25) deg, 
and other stars, some of which concentrated along the bow shape  pattern formed by the dust,
limited by the SFO\,55 BRCs and the reflection nebula NGC\,2626,
and a second small group of stars, with an elongated distribution, 
around  (RA, Dec)=(129.87, -41.17). 
The second generation of stars formed  two other small clumps of stars  in 
proximity of the O8.5IV star HD73882, a prominent clump around the reflection nebula NGC\,2626
and a small clump around SFO\,55.
More recent SF occured around NGC\,2626,
around  (RA, Dec)=(129.87, -41.17) and 
again,  along the bow-shape  pattern formed by the dust around (RA, Dec)=(129.7, -40.65),
going up along the filament formed by the dust.
 In this region, the difference between the spatial distributions of the first and the second generation
of stars is marginally significant ({\it p}-value=0.04) but the difference between the  spatial distributions
of the youngest stars and the other groups is significant with {\it p}-value=0.00006 in both cases. 

\subsubsection{Concluding remarks} 
 The spatial distribution of the M-type members suggests the formation of several clumps 
 correlated to the stellar ages revealing that SF in young 
 clusters occurs  in small subclusters, as  predicted by models \citep{clar00}.

For the YSOs selected with the classical methods we are not able to derive accurate individual reddenings
and therefore ages. However,  the spatial distribution of these objects is consistent
with a population formed as the M-type selected members. The spatial distribution of
 most  YSOs in RCW\,33  is consistent with that found for the first generation of M-type stars,
  while that of  YSOs in RCW\,32 and RCW\,27 is
 better consistent with the one found for the second generation of M-type stars.

Our results suggest a significant age spread with 
evidence of a continous SF  process that started about 12-13\,Myr ago and lasted
 until less than 1\,Myr
ago. The age distribution in the three regions is consistent with  a decreasing SF rate
only in Vela T2 (RCW\,33), while the age distribution of  YSOs in Cr\,197 (RCW\,32) 
might suggest the presence of a triggering process likely due to the ionizing star HD\,74804,
 that is consistent with the
scenario suggested by \citet{urqu09} in which the \hii region radiation causes the compression and the
 fragmentation of the material in the molecular cloud.
 
A strong evidence of triggered SF  has also been found in Vela T1 (RCW\,27), where several clumps 
formed with  time. The presence of the two peaks  in the age distribution suggests a SF
acceleration likely caused by the physical interaction between the 
 radiation of massive stars and the surrounding molecular cloud. 


 To estimate the total number of stars and the total mass of the three populations,
we assume a universal Initial Mass Function (IMF) and
consider  the completeness  limits  of the magnitudes for  M-type stars.
that are   r=19.97 ,   i= 19.09,     J=15.87  ,   H=15.09   and    K=14.4250.
These values have been estimated
at  the peak of the magnitude distribution. As described in \citet{dami18},
the minimum mass detected at a given distance strongly depends on the age and the absorption
A$_V$ \citep[see Fig.\,5 of ][]{dami18}. 
 To estimate the lowest mass above which our sample is complete,
 we used two isochrones at  two representative  ages for the three regions, i.e. 4 and 5\,Myr,
and the median A$_V$ of the three regions that are A$_V^{\rm med}$=[1.4,2.5,1.8] respectively for RCW\,33, 
 RCW\,32 and RCW\,27. 
Under these conditions, we estimated that the mass ranges in which  M-type stars younger than 4 and 5\,Myr
and 
with A$_V<$A$_V^{\rm med}$ are all detected
 are 
[0.22-0.94]\,M$_\odot$, [0.34-0.94]\,M$_\odot$, [0.27-0.94]\,M$_\odot$ for 4\,Myr and
[0.24-0.94]\,M$_\odot$, [0.36-0.94]\,M$_\odot$, [0.28-0.94]\,M$_\odot$ for 5\,Myr, respectively,
for RCW\,33, RCW\,32 and RCW\,27.
The upper bound of the mass ranges were derived from 
 the 4 and 5\,Myr isochrones for i-K$>$2.2,
that is the color limit used for the selection of M-type stars (see Fig.\,13).
 
In these ranges, oldest and more reddened objects cannot be detected.
For each of the three clusters, we counted the number of stars within these mass ranges, with 
 A$_V$ smaller than the median value of each region and ages smaller than
4 (5\,Myr) that are 30, 18 and 62 (42, 24 and  64), respectively in RCW\,33,  RCW\,32 and RCW\,27.
To  account for the objects in the same mass ranges, but with ages older
than 4 and 5\,Myr,  
 we used  the fraction of 
M-type stars in these mass ranges and 
within the same limits on A$_V$
but with ages older than 4 (5\,Myr),
that are 0.70,    0.69 and 0.53 (0.58,     0.55 and     0.47).
Finally, we extrapolated the total number of stars of the three populations 
across the entire mass spectrum by considering the 
fraction of stars predicted by the \citet{weid10} IMF in the same mass ranges that are
    0.34, 0.21 and     0.28 (0.32,    0.20 and   0.27). 
    We thus estimate  that the total number of YSOs for the three populations are 
    in the range [124-224], [122-216] and [403-493] stars, respectively,  in RCW\,33,  RCW\,32 and RCW\,27.
   If we consider the total mass of M-type stars within the previous mass ranges and consider the
   mass fraction predicted in the same ranges with  the \citet{weid10} IMF, we estimate that the total mass
   of the three populations is [50-86]\,M$_\odot$, [45-81]\,M$_\odot$ and [141-174]\,M$_\odot$.
   The large uncertainties on these estimates are mainly due to the limited depth of the 2MASS photometry,
   that hopefully will be overcome by future deep IR surveys covering this region. We note that 
   by using the limit of 4\,Myr, the  mass ranges are slightly  larger than by using  5\,Myr and this
   allows us a slightly less uncertain extrapolation of the IMF, but at the same time we include
   a smaller fraction of objects from our computation and consequently a larger correction for older stars
    is necessary. Thus, our  choice  to consider both two 4 and 5\,Myr isochrones enable us to
    obtain an estimate as reliable as possible, that, however should be considered with caution and,
    it is likely a lower limit since
    we did not apply a correction for  YSOs with A$_V>$A$_V^{\rm med}$.

\section{Summary and conclusions}
We used  optical and NIR photometric surveys to study the young stellar population
in the three \hii regions RCW\,33, 32 and 27 where \citet{pett94} found signatures 
of recent SF  from H$\alpha$ observations.
Using the deeper VPHAS+ survey that covers the region studied by \citet{pett94},
we found 329 YSOs with H$\alpha$ emission showing a spatial pattern that confirms the
presence of young clusters associated to the three \hii regions.
This result prompted the search of the associated population of YSOs surrounded by 
a circumstellar disk, through  NIR surveys, such as SPITZER/GLIMPSE and 2MASS.
Using the known properties of YSOs with disk, we selected 559 class\,II and 20 class\,I YSOs.
We exploited the recent Gaia DR1 to study the kinematic population of the three regions and
we found  10, 19 and 25 stars sharing similar PMs and parallaxes in Cr\,197, Vela T2 and Vela T1, respectively,
confirming the presence of the three clusters, and with 
a placement in  the CMD  consistent with  PMS 
stars   at the distance of about 750\,pc. The distance of the three regions
derived from the Gaia data is consistent with that derived from the CMD of the low mass population.
Additional YSOs were found by using also archive ROSAT and Chandra X-ray observations. 
At the end, we selected a total of 907 candidate YSOs in the whole region,
with  different spatial distribution patterns in the three \hii regions.

We added, to this sample,  a further sample of  1\,200 candidate M-type members.
These objects were selected by exploiting two important features of M-type stars,
that are the significantly higher luminosity in the PMS phase with respect to MS stars of the
same spectral type and the sensitivity of  their  i-J, i-H and i-K colors to stellar gravity
that allowed us to derive the individual reddening indipendently  in three different CCDs, and to
select the best A$_V$ value for each M-type star. 
For the whole M-type population, including the field star population,
 we  therefore  derived the unreddened CMD where the only unknown  is the distance.
  
The comparison of the unreddened CMD with  PMS isochrones, at the distance
of the clusters, allowed us to discard the 
foreground population of M-type MS stars and the 
background  population of M-type giants. In addition, based on the spatial analysis,  
we found that  the populations of YSOs in RCW\,33, 32 and 27
are dominated by objects with A$_V>$1 and this allowed us to discard also the low reddened
(A$_V<$1) population of M-type MS stars contaminating the photometric PMS locus
in the CMD of the clusters.

The spatial distribution of the M-type YSOs is strongly correlated to that of the YSOs selected with classical
methods. Since we have a subsample of 72
 M-type stars selected with both methods,
we find a total population of  2\,035 
 YSOs, mainly distributed inside the three \hii regions,
presenting for the first time, a stellar census of YSOs in NW-VMR, down to the M-type stars.
For the M-type YSO subsample, we also derived stellar masses, down to 0.1\,M$_\odot$, and stellar
ages in the range 0.5-13\,Myr.

We performed a detailed spatial analysis of the M-type members in the three \hii regions
and compared them with  WISE images at 12\,$\mu$m, that highlights the regions with high concentration
of dust. By splitting the M-type YSOs in three different ranges of ages, we found indications
on the SF rate. In particular, we found that the formation of YSOs in  clumps  or 
in elongated features, is connected to the ages and to the presence of BRCs,
visible in the WISE image.

The most likely scenario is that,
 while the SF rate in RCW\,33 is decreasing with time, 
the SF process in RCW\,32 and 27 has been likely triggered by the interaction
between the \hii radiation associated to the ionizing stars and the molecular clouds
from which the low mass star population formed. This scenario is consistent with that
suggested by \citet{urqu09}, based on CO and mid-IR observations of a sample of \hii regions,
including RCW\,32 and RCW\,27.

The criteria used in this work to select M-type stars can be exploited in future works
in several contexts of stellar astrophysics. In the context of SF, 
the intrinsic higher luminosity of M-type stars in PMS phase  with respect to MS stars
is a new observational diagnostics that allows  detection of the M-type young populations
of relatively distant regions ($\sim$1-2\,kpc) by using 
 recent photometric surveys. The M-type population in these regions has been
up to now mainly revealed by deep and high sensitive X-ray or spectroscopic observations,  that are
significantly more expensive in terms of  telescope time demand. In addition,
 due to the limited FOV, it is very difficult, if not impossible, to study large fields
 of several square degrees, as, instead, can be easily done with the current photometric surveys.

The Gaia DR2  data will likely allow a  more refined selections
 since, with the kinematic parameters, it will be possible to confirm
 M-type candidate members. Moreover, other available and future very deep photometric surveys, such as 
for example PanSTARRS, UKIDSS, WISE, VVV and  LSST, will allow us to detect even more distant 
M-type stars also in even more distant star forming regions. 
 
Finally, the opportunity to derive accurate individual reddening values of M-type stars
is a new tool that can be exploited to discard M-type contaminants, both MS or giants, and
therefore to select M-type members, based on the known parameters of the clusters we are dealing with.
Moreover, this new diagnostic can be exploited to  
distinguish  M-type MS stars from giants in fields
where YSOs are not expected to be found. 

\begin{acknowledgements}
We wish to thank the anonymous referee for his/her  interesting comments and
 suggestions.
This work is based on observations made with ESO Telescopes at the La Silla or Paranal 
Observatories under programme ID(s) 177.D-3023(B), 177.D-3023(C), 177.D-3023(D), 177.D-3023(E).
This work has made use of data from the European Space Agency (ESA)
mission {\it Gaia} (\url{http://www.cosmos.esa.int/gaia}), processed by
the {\it Gaia} Data Processing and Analysis Consortium (DPAC,
\url{http://www.cosmos.esa.int/web/gaia/dpac/consortium}). Funding
for the DPAC has been provided by national institutions, in particular
the institutions participating in the {\it Gaia} Multilateral Agreement.
This publication makes use of data products from the Wide-field Infrared 
Survey Explorer, which is a joint project of the University of California,
 Los Angeles, and the Jet Propulsion Laboratory/California Institute of Technology, 
 funded by the National Aeronautics and Space Administration.
\end{acknowledgements}
\bibliographystyle{aa}
\bibliography{/Users/prisinzano/BIBLIOGRAPHY/bibdesk}

\begin{thebibliography}{48}
\expandafter\ifx\csname natexlab\endcsname\relax\def\natexlab#1{#1}\fi

\bibitem[{{Benjamin} {et~al.}(2003){Benjamin}, {Churchwell}, {Babler}, {Bania},
  {Clemens}, {Cohen}, {Dickey}, {Indebetouw}, {Jackson}, {Kobulnicky},
  {Lazarian}, {Marston}, {Mathis}, {Meade}, {Seager}, {Stolovy}, {Watson},
  {Whitney}, {Wolff}, \& {Wolfire}}]{benj03}
{Benjamin}, R.~A., {Churchwell}, E., {Babler}, B.~L., {et~al.} 2003, \pasp,
  115, 953

\bibitem[{{Bonatto} \& {Bica}(2010)}]{bona10}
{Bonatto}, C. \& {Bica}, E. 2010, \aap, 516, A81

\bibitem[{{Cardelli} {et~al.}(1989){Cardelli}, {Clayton}, \& {Mathis}}]{card89}
{Cardelli}, J.~A., {Clayton}, G.~C., \& {Mathis}, J.~S. 1989, \apj, 345, 245

\bibitem[{{Churchwell} {et~al.}(2009){Churchwell}, {Babler}, {Meade},
  {Whitney}, {Benjamin}, {Indebetouw}, {Cyganowski}, {Robitaille}, {Povich},
  {Watson}, \& {Bracker}}]{chur09}
{Churchwell}, E., {Babler}, B.~L., {Meade}, M.~R., {et~al.} 2009, \pasp, 121,
  213

\bibitem[{{Cieza} {et~al.}(2005){Cieza}, {Kessler-Silacci}, {Jaffe}, {Harvey},
  \& {Evans}}]{ciez05}
{Cieza}, L.~A., {Kessler-Silacci}, J.~E., {Jaffe}, D.~T., {Harvey}, P.~M., \&
  {Evans}, II, N.~J. 2005, \apj, 635, 422

\bibitem[{{Clarke} {et~al.}(2000){Clarke}, {Bonnell}, \&
  {Hillenbrand}}]{clar00}
{Clarke}, C.~J., {Bonnell}, I.~A., \& {Hillenbrand}, L.~A. 2000, Protostars and
  Planets IV, 151

\bibitem[{{Crampton} \& {Fisher}(1974)}]{cram74}
{Crampton}, D. \& {Fisher}, W.~A. 1974, Publications of the Dominion
  Astrophysical Observatory Victoria, 14, 283

\bibitem[{{Cutri} {et~al.}(2003){Cutri}, {Skrutskie}, {van Dyk}, {Beichman},
  {Carpenter}, {Chester}, {Cambresy}, {Evans}, {Fowler}, {Gizis}, {Howard},
  {Huchra}, {Jarrett}, {Kopan}, {Kirkpatrick}, {Light}, {Marsh}, {McCallon},
  {Schneider}, {Stiening}, {Sykes}, {Weinberg}, {Wheaton}, {Wheelock}, \&
  {Zacarias}}]{cutr03}
{Cutri}, R.~M., {Skrutskie}, M.~F., {van Dyk}, S., {et~al.} 2003, {2MASS All
  Sky Catalog of point sources.} (The IRSA 2MASS All-Sky Point Source Catalog,
  NASA/IPAC Infrared Science
  Archive.~http://irsa.ipac.caltech.edu/applications/Gator/)

\bibitem[{{Damiani}(2018)}]{dami18}
{Damiani}, F. 2018, ArXiv e-prints

\bibitem[{{Damiani} {et~al.}(2017){Damiani}, {Pillitteri}, \&
  {Prisinzano}}]{dami17}
{Damiani}, F., {Pillitteri}, I., \& {Prisinzano}, L. 2017, \aap, 602, A115

\bibitem[{{de Zeeuw} {et~al.}(1999){de Zeeuw}, {Hoogerwerf}, {de Bruijne},
  {Brown}, \& {Blaauw}}]{de-z99}
{de Zeeuw}, P.~T., {Hoogerwerf}, R., {de Bruijne}, J.~H.~J., {Brown}, A.~G.~A.,
  \& {Blaauw}, A. 1999, \aj, 117, 354

\bibitem[{{Dias} {et~al.}(2002){Dias}, {Alessi}, {Moitinho}, \&
  {L{\'e}pine}}]{dias02}
{Dias}, W.~S., {Alessi}, B.~S., {Moitinho}, A., \& {L{\'e}pine}, J.~R.~D. 2002,
  \aap, 389, 871

\bibitem[{{Drew} {et~al.}(2014){Drew}, {Gonzalez-Solares}, {Greimel}, {Irwin},
  {K{\"u}pc{\"u} Yoldas}, {Lewis}, {Barentsen}, {Eisl{\"o}ffel}, {Farnhill},
  {Martin}, {Walsh}, {Walton}, {Mohr-Smith}, {Raddi}, {Sale}, {Wright},
  {Groot}, {Barlow}, {Corradi}, {Drake}, {Fabregat}, {Frew}, {G{\"a}nsicke},
  {Knigge}, {Mampaso}, {Morris}, {Naylor}, {Parker}, {Phillipps}, {Ruhland},
  {Steeghs}, {Unruh}, {Vink}, {Wesson}, \& {Zijlstra}}]{drew14}
{Drew}, J.~E., {Gonzalez-Solares}, E., {Greimel}, R., {et~al.} 2014, \mnras,
  440, 2036

\bibitem[{{Fasano} \& {Franceschini}(1987)}]{fasa87}
{Fasano}, G. \& {Franceschini}, A. 1987, \mnras, 225, 155

\bibitem[{{Gaia Collaboration} {et~al.}(2016{\natexlab{a}}){Gaia
  Collaboration}, {Brown}, {Vallenari}, {Prusti}, {de Bruijne}, {Mignard},
  {Drimmel}, {Babusiaux}, {Bailer-Jones}, {Bastian}, \& et~al.}]{gaia16}
{Gaia Collaboration}, {Brown}, A.~G.~A., {Vallenari}, A., {et~al.}
  2016{\natexlab{a}}, \aap, 595, A2

\bibitem[{{Gaia Collaboration} {et~al.}(2016{\natexlab{b}}){Gaia
  Collaboration}, {Prusti}, {de Bruijne}, {Brown}, {Vallenari}, {Babusiaux},
  {Bailer-Jones}, {Bastian}, {Biermann}, {Evans}, \& et~al.}]{gaia16a}
{Gaia Collaboration}, {Prusti}, T., {de Bruijne}, J.~H.~J., {et~al.}
  2016{\natexlab{b}}, \aap, 595, A1

\bibitem[{{Gaustad} {et~al.}(2001){Gaustad}, {McCullough}, {Rosing}, \& {Van
  Buren}}]{gaus01}
{Gaustad}, J.~E., {McCullough}, P.~R., {Rosing}, W., \& {Van Buren}, D. 2001,
  \pasp, 113, 1326

\bibitem[{{Georgelin} {et~al.}(1973){Georgelin}, {Georgelin}, \&
  {Roux}}]{geor73}
{Georgelin}, Y.~M., {Georgelin}, Y.~P., \& {Roux}, S. 1973, \aap, 25, 337

\bibitem[{{Gutermuth} {et~al.}(2009){Gutermuth}, {Megeath}, {Myers}, {Allen},
  {Pipher}, \& {Fazio}}]{gute09}
{Gutermuth}, R.~A., {Megeath}, S.~T., {Myers}, P.~C., {et~al.} 2009, \apjs,
  184, 18

\bibitem[{{Henden} {et~al.}(2016){Henden}, {Templeton}, {Terrell}, {Smith},
  {Levine}, \& {Welch}}]{hend16}
{Henden}, A.~A., {Templeton}, M., {Terrell}, D., {et~al.} 2016, VizieR Online
  Data Catalog, 2336

\bibitem[{{Indebetouw} {et~al.}(2005){Indebetouw}, {Mathis}, {Babler}, {Meade},
  {Watson}, {Whitney}, {Wolff}, {Wolfire}, {Cohen}, {Bania}, {Benjamin},
  {Clemens}, {Dickey}, {Jackson}, {Kobulnicky}, {Marston}, {Mercer},
  {Stauffer}, {Stolovy}, \& {Churchwell}}]{inde05}
{Indebetouw}, R., {Mathis}, J.~S., {Babler}, B.~L., {et~al.} 2005, \apj, 619,
  931

\bibitem[{{Kenyon} \& {Hartmann}(1995)}]{keny95}
{Kenyon}, S.~J. \& {Hartmann}, L. 1995, \apjs, 101, 117

\bibitem[{{Kharchenko} {et~al.}(2005){Kharchenko}, {Piskunov}, {R{\"o}ser},
  {Schilbach}, \& {Scholz}}]{khar05}
{Kharchenko}, N.~V., {Piskunov}, A.~E., {R{\"o}ser}, S., {Schilbach}, E., \&
  {Scholz}, R.-D. 2005, \aap, 440, 403

\bibitem[{{Kharchenko} {et~al.}(2013){Kharchenko}, {Piskunov}, {Schilbach},
  {R{\"o}ser}, \& {Scholz}}]{khar13}
{Kharchenko}, N.~V., {Piskunov}, A.~E., {Schilbach}, E., {R{\"o}ser}, S., \&
  {Scholz}, R.-D. 2013, \aap, 558, A53

\bibitem[{{Liseau} {et~al.}(1992){Liseau}, {Lorenzetti}, {Nisini}, {Spinoglio},
  \& {Moneti}}]{lise92}
{Liseau}, R., {Lorenzetti}, D., {Nisini}, B., {Spinoglio}, L., \& {Moneti}, A.
  1992, \aap, 265, 577

\bibitem[{{Lorenzetti} {et~al.}(1993){Lorenzetti}, {Spinoglio}, \&
  {Liseau}}]{lore93}
{Lorenzetti}, D., {Spinoglio}, L., \& {Liseau}, R. 1993, \aap, 275, 489

\bibitem[{{Marigo} {et~al.}(2017){Marigo}, {Girardi}, {Bressan}, {Rosenfield},
  {Aringer}, {Chen}, {Dussin}, {Nanni}, {Pastorelli}, {Rodrigues}, {Trabucchi},
  {Bladh}, {Dalcanton}, {Groenewegen}, {Montalb{\'a}n}, \& {Wood}}]{mari17}
{Marigo}, P., {Girardi}, L., {Bressan}, A., {et~al.} 2017, \apj, 835, 77

\bibitem[{{Massi} {et~al.}(1999){Massi}, {Giannini}, {Lorenzetti}, {Liseau},
  {Moneti}, \& {Andreani}}]{mass99}
{Massi}, F., {Giannini}, T., {Lorenzetti}, D., {et~al.} 1999, \aaps, 136, 471

\bibitem[{{Massi} {et~al.}(2003){Massi}, {Lorenzetti}, \& {Giannini}}]{mass03}
{Massi}, F., {Lorenzetti}, D., \& {Giannini}, T. 2003, \aap, 399, 147

\bibitem[{{Massi} {et~al.}(2006){Massi}, {Testi}, \& {Vanzi}}]{mass06}
{Massi}, F., {Testi}, L., \& {Vanzi}, L. 2006, \aap, 448, 1007

\bibitem[{{Meyer} {et~al.}(1997){Meyer}, {Calvet}, \& {Hillenbrand}}]{meye97}
{Meyer}, M.~R., {Calvet}, N., \& {Hillenbrand}, L.~A. 1997, \aj, 114, 288

\bibitem[{{Murphy} \& {May}(1991)}]{murp91}
{Murphy}, D.~C. \& {May}, J. 1991, \aap, 247, 202

\bibitem[{{O'Donnell}(1994)}]{odon94}
{O'Donnell}, J.~E. 1994, \apj, 422, 158

\bibitem[{{Oke} \& {Gunn}(1983)}]{oke83}
{Oke}, J.~B. \& {Gunn}, J.~E. 1983, \apj, 266, 713

\bibitem[{{Olmi} {et~al.}(2009){Olmi}, {Ade}, {Angl{\'e}s-Alc{\'a}zar}, {Bock},
  {Chapin}, {De Luca}, {Devlin}, {Dicker}, {Elia}, {Fazio}, {Giannini},
  {Griffin}, {Gundersen}, {Halpern}, {Hargrave}, {Hughes}, {Klein},
  {Lorenzetti}, {Marengo}, {Marsden}, {Martin}, {Massi}, {Mauskopf},
  {Netterfield}, {Patanchon}, {Rex}, {Salama}, {Scott}, {Semisch}, {Smith},
  {Strafella}, {Thomas}, {Truch}, {Tucker}, {Tucker}, {Viero}, \&
  {Wiebe}}]{olmi09}
{Olmi}, L., {Ade}, P.~A.~R., {Angl{\'e}s-Alc{\'a}zar}, D., {et~al.} 2009, \apj,
  707, 1836

\bibitem[{{Pettersson}(2008)}]{pett08}
{Pettersson}, B. 2008, {Young Stars and Dust Clouds in Puppis and Vela}, ed.
  B.~{Reipurth}, 43

\bibitem[{{Pettersson} \& {Reipurth}(1994)}]{pett94}
{Pettersson}, B. \& {Reipurth}, B. 1994, \aaps, 104

\bibitem[{{Randich} {et~al.}(2017){Randich}, {Tognelli}, {Jackson}, {Jeffries},
  {Degl'Innocenti}, {Pancino}, {Re Fiorentin}, {Spagna}, {Sacco}, {Bragaglia},
  {Magrini}, {Prada Moroni}, {Alfaro}, {Franciosini}, {Morbidelli},
  {Roccatagliata}, {Bouy}, {Bravi}, {Jim{\'e}nez-Esteban}, {Jordi}, {Zari},
  {Tautvai{\v s}iene}, {Drazdauskas}, {Mikolaitis}, {Gilmore}, {Feltzing},
  {Vallenari}, {Bensby}, {Koposov}, {Korn}, {Lanzafame}, {Smiljanic}, {Bayo},
  {Carraro}, {Costado}, {Heiter}, {Hourihane}, {Jofr{\'e}}, {Lewis}, {Monaco},
  {Prisinzano}, {Sbordone}, {Sousa}, {Worley}, \& {Zaggia}}]{rand17}
{Randich}, S., {Tognelli}, E., {Jackson}, R., {et~al.} 2017, ArXiv e-prints

\bibitem[{{Stetson}(1987)}]{stet87}
{Stetson}, P.~B. 1987, \pasp, 99, 191

\bibitem[{{Sugitani} \& {Ogura}(1994)}]{sugi94}
{Sugitani}, K. \& {Ogura}, K. 1994, \apjs, 92, 163

\bibitem[{{Testi} {et~al.}(2001){Testi}, {Vanzi}, \& {Massi}}]{test01}
{Testi}, L., {Vanzi}, L., \& {Massi}, F. 2001, The Messenger, 103, 28

\bibitem[{{Tognelli} {et~al.}(2011){Tognelli}, {Prada Moroni}, \&
  {Degl'Innocenti}}]{togn11}
{Tognelli}, E., {Prada Moroni}, P.~G., \& {Degl'Innocenti}, S. 2011, \aap, 533,
  A109

\bibitem[{{Urquhart} {et~al.}(2009){Urquhart}, {Morgan}, \&
  {Thompson}}]{urqu09}
{Urquhart}, J.~S., {Morgan}, L.~K., \& {Thompson}, M.~A. 2009, \aap, 497, 789

\bibitem[{{Wang} {et~al.}(2016){Wang}, {Liu}, {Qiu}, {Bai}, {Yang}, {Guo}, \&
  {Zhang}}]{wang16}
{Wang}, S., {Liu}, J., {Qiu}, Y., {et~al.} 2016, \apjs, 224, 40

\bibitem[{{Weidner} {et~al.}(2010){Weidner}, {Kroupa}, \& {Bonnell}}]{weid10}
{Weidner}, C., {Kroupa}, P., \& {Bonnell}, I.~A.~D. 2010, \mnras, 401, 275

\bibitem[{{Yamaguchi} {et~al.}(1999{\natexlab{a}}){Yamaguchi}, {Mizuno},
  {Saito}, {Matsunaga}, {Mizuno}, {Ogawa}, \& {Fukui}}]{yama99}
{Yamaguchi}, N., {Mizuno}, N., {Saito}, H., {et~al.} 1999{\natexlab{a}}, \pasj,
  51, 775

\bibitem[{{Yamaguchi} {et~al.}(1999{\natexlab{b}}){Yamaguchi}, {Saito},
  {Mizuno}, {Mine}, {Mizuno}, {Ogawa}, \& {Fukui}}]{yama99a}
{Yamaguchi}, R., {Saito}, H., {Mizuno}, N., {et~al.} 1999{\natexlab{b}}, \pasj,
  51, 791

\bibitem[{{Zasowski} {et~al.}(2009){Zasowski}, {Majewski}, {Indebetouw},
  {Meade}, {Nidever}, {Patterson}, {Babler}, {Skrutskie}, {Watson}, {Whitney},
  \& {Churchwell}}]{zaso09}
{Zasowski}, G., {Majewski}, S.~R., {Indebetouw}, R., {et~al.} 2009, \apj, 707,
  510

\end{thebibliography}

\end{document}